\documentclass[10pt,journal,compsoc]{IEEEtran}
\usepackage{dsfont}
\usepackage{threeparttable}
\usepackage[nocompress]{cite}
\usepackage{cite}
\usepackage{indentfirst}
\usepackage{amsmath}
\usepackage{amsfonts}
\usepackage{color}
\usepackage{comment}
\usepackage{graphicx}
\usepackage{multirow}
\usepackage{algorithmic}
\usepackage{bm}
\usepackage{url}
\usepackage{booktabs}  
\usepackage{bbding}
\usepackage{enumitem}
\usepackage{float}
\usepackage{array}
\usepackage{caption}
\usepackage{float}
\usepackage{subfig}
\makeatletter
\newcommand{\thickhline}{%
    \noalign {\ifnum 0=`}\fi \hrule height 1pt
    \futurelet \reserved@a \@xhline
}

\begin{document}

\title{ Disentangled Graph Neural Networks for Session-based Recommendation }

\author{Ansong Li,
        Zhiyong Cheng,
        Fan Liu,
        Zan Gao,
        Weili Guan,
        Yuxin Peng, ~{Senior Member, IEEE}

 \IEEEcompsocitemizethanks{
\IEEEcompsocthanksitem A. Li is with Shandong Artificial Intelligence Institute, Qilu University of Technology (Shandong Academy of Sciences), and also with School of Software, Xi'an Jiaotong University.\protect  E-mail: lasnling@gmail.com. 
\IEEEcompsocthanksitem The work was done when A. Li was an intern at Shandong Artificial Intelligence Institute, Qilu University of Technology (Shandong Academy of
Sciences).\\
\IEEEcompsocthanksitem Z. Cheng and Z. Gao are with  Shandong Artificial Intelligence Institute, Qilu University of Technology (Shandong Academy of Sciences), China. Email: jason.zy.cheng@gmail.com. Z. Cheng is the  corresponding author.\\
\IEEEcompsocthanksitem F. Liu is with School of Computing, National University of Signapore. Email: liufancs@gmail.com.
\IEEEcompsocthanksitem W. Guan is with Monash University, Australia. Email: weili.guan@monash.edu.\\
\IEEEcompsocthanksitem Y. Peng is with Wangxuan Institute of Computer Technology, Peking University, Beijing 100871, China (e-mail: pengyuxin@pku.edu.cn).
}
}

\markboth{IEEE TRANSACTIONS ON KNOWLEDGE AND DATA ENGINEERING}%
{Shell \MakeLowercase{\textit{et al..}}: Bare Demo of IEEEtran.cls for Computer Society Journals}

\IEEEtitleabstractindextext{%
\begin{abstract}
  Session-based recommendation (SBR) has drawn increasingly research attention in recent years, due to its great practical value by only exploiting the limited user behavior history in the current session. The key of SBR is to accurately infer the anonymous user purpose in a session which is typically represented as session embedding, and then match it with the item embeddings for the next item prediction.  Existing methods typically learn the session embedding at the item level, namely, aggregating the embeddings of items with or without the attention weights assigned to items. However, they ignore the fact that a user's intent on adopting an item is driven by certain factors of the item (e.g., the \emph{leading actors} of an movie). In other words, they have not explored finer-granularity interests of users at the factor level to generate the session embedding, leading to sub-optimal performance. To address the problem, we propose a novel method called Disentangled Graph Neural Network (Disen-GNN) to capture the session purpose with the consideration of factor-level attention on each item. Specifically, we first employ the disentangled learning technique to cast item embeddings into the embedding of multiple factors, and then use the gated graph neural network (GGNN) to learn the embedding factor-wisely based on the item adjacent similarity matrix computed for each factor. Moreover, the distance correlation is adopted to enhance the independence between each pair of factors. After representing each item with independent factors, an attention mechanism is designed to learn user intent to different factors of each item in the session. The session embedding is then generated by aggregating the item embeddings with attention weights of each item's factors. To this end, our model takes user intents at the factor level into account to infer the user purpose in a session. Extensive experiments on three benchmark datasets demonstrate the superiority of our method over existing methods. 
  
\end{abstract}

}

\maketitle

\IEEEdisplaynontitleabstractindextext

\IEEEpeerreviewmaketitle

\setlength{\parindent}{2em}
\vspace{-0.6cm}
\section{Introduction} \label{sec:introduction}

\IEEEPARstart{R}{ecommendation} system plays an increasingly important role in assisting people in finding their desired information. With decades of development,  a variety of recommendation algorithms have been developed and deployed on various platforms to provide recommendation services, such as E-commerce websites and streaming media provider. Typical recommendation methods, such as content-based recommendation \cite{CB-RS}\cite{F-CB}, collaborative filtering \cite{NGCF}\cite{Light-GCN} and hybrid recommender systems \cite{Hybrid}, have achieved great success by directly or indirectly exploit user profiles like clicking or purchasing behaviors to model user preference for accurate recommendation. However, users' identity information is often unavailable in many real-world scenarios, e.g., unregistered users or the ones who are reluctant to log in for privacy concerns. In such situations,  only the behavior history on the current session can be leveraged. Thereby, there is a urgently practical need to provide accurate recommendations with the limited behavior information. Aiming to predict the next interested item for a give anonymous behavior sequence in chronological order is the so-called session-based recommendation (SBR), which has attracted increasingly research interests in recent years.

The early SBR methods mainly based on the similarity-based~\cite{ITEM-KNN} or Markov chain-based method~\cite{FPMC}. The former methods make recommendations based on the co-occurrence patterns of items while ignoring the sequential information, and the latter ones have a strong sequential assumption that the next item is solely  based on the previous one,   failing to capture long-term dependence. The emergence and fast development of deep learning techniques provide a solution to alleviate the problem, especially the ones for modeling sequential data, such as recurrent neural network (RNN) and graph neural network (GNN).  Hidasi et al.\cite{SR-RNN} made the first attempt to apply RNN  in SBR and proposed the   GRU4REC model, which captures the transition relationship with gated recurrent unite (GRU). Later on, more advanced neural network based methods have been developed, such as combining RNN with attention network (NARM)~\cite{NARM} or memory network (STAMP)~\cite{STAMP}. By constructing an item graph for each session, GNN-based methods have also been applied to capture the transition relations between distant items by learning item embedding via information propagation and updating over the graph. For example, Wu et al.\cite{SR-GNN} proposed SR-GNN~\cite{SR-GNN} to model the higher-order item transitions with GNN. With the success of SR-GNN, more GNN-based models have been proposed, such as the ones using attention mechanism~\cite{GC-SAN,FGNN,TAGNN} and global information of all sessions\cite{FGNN-BCS,GCE-GNN}.


In SBR, the prediction is made by matching the representation of the target item with \emph{the main purpose of the session}, which is typically represented as \emph{the session embedding}. Therefore, learning good representations for items and sessions is crucial for accurate recommendation. The representation of a session is often obtained by aggregating the representations of all items within this session. Since there could be irrelevant items and the user intent can be changed along with the items in a session,  some methods have been proposed to assign different weights to items~\cite{NARM,FGNN} or place more weight to the last item  in a session when generating the session representation~\cite{STAMP,SR-GNN}. 
Despite the great progress has been achieved thus far, we argue that existing methods have not distinguished the importance of different factors of the items in a session when modeling the main purpose of a session.  It is well-recognized that a user's intents are diverse when adopting items and  an item are characterized by various factors (e.g.,  \emph{color}, \emph{style}, \emph{brand} for \emph{clothes}, and \emph{actor}, \emph{director}, \emph{plot} for \emph{movies} )~\cite{cheng20183ncf}. The intent of a user on adopting an item is driven by the factors that she is
interested in. It is relatively easy to know a user's interest in  which type of items (e.g., clothes or movie) in a session. To make accurate prediction to the next item, inferring \emph{which factors the users pay attention to in previous items and their features} becomes more important. For example, is the \emph{actor} or \emph{plot} the key to recommend the next movie? And what kinds of plots attract the user most? However, it is challenging to infer the factors that the user cares most, especially considering the different items and different prominent factors of each item in a session.

Existing methods in SBR often represent each item with a holistic representation (or embedding), which does not separate the features of different factors. As a result, it cannot identify the key factors when modeling the main purpose of a session. To tackle this problem, we propose a novel Disentangled Graph Neural Network model (Disen-GNN) for SBR in this paper.  In our model, the item embedding is cast into a few chunks with the assumption that each chunk represents the embedding of a latent factor. Moreover, the factor embeddings are learned separately through embedding propagation based on a factor-based similarity matrix over the session graph using GGNN. And the distance correlation is adopted to further enhance the independence between every pair of factors. In this way, our model represents each item as the embeddings of independent factors. Subsequently, an attention mechanism is designed to assign weights to different factors of each item in the session, and then the session embedding is obtained by aggregating the embedding of each factor with the assigned weights across all the session items. To this end, Disen-GNN models the main purpose of a session (i.e., session embedding) by considering the user attentions to different factors of all items in the session. Finally, the recommendation is made based on the aggregation of the factor-wise similarities between the embeddings of the session and the target item. 

To evaluate the performance of Disen-GNN, we perform extensively experiments on three public datasets. Experimental results show that our model can achieve remarkable improvement over a variety of strong competitors, including the recently proposed GNN-based models. Further ablation studies demonstrate the potential of using disentangled representation learning in SBR and validate the effectiveness of different components in our model. In summary, the main contributions of this work are threefold:
\begin{itemize}[leftmargin=*]
  \item We highlight the importance of considering the different contributions of item factors to capture the user purpose in a session for SBR. As far as we known, it is the first to consider the factor-level attention to different items and model session embedding at the factor-level for SBR.
  \item We propose a novel disentangled GNN model for SBR, which represents item embeddings with disentangled representations of factors and generates session embeddings by aggregating item embeddings with assigned factor-level attention weights. Our model also sheds light on the potential of applying disentangled representation learning techniques in SBR.
  \item We have conducted extensive experiments on three real-world datasets to evaluate the effectiveness of the proposed Disen-GNN model and performed ablation study to examine the validity of different components in our model. Experiments demonstrate the superiority of our model over the state-of-the-art methods.
\end{itemize}

\section{Related Work}


\subsection{Session-based Recommendation}
 Session-based recommendation (SBR) has attracted increasingly research attentions and many approaches have been proposed in recent years. SBR is to predict the next item based on the sequence of previous items. The Markov chain-based methods, which infers a user's next action based on the previous one, can be naturally adapted for SBR. For example, the FPMC\cite{FPMC} method, which combines matrix factorization and the first-order Markov chain to model the sequential pattern and  user preference for recommendation, can be adapted for SBR by ignoring the latent user representation since there is no user information in the anonymous sessions. The drawback of Markov chain-based methods is that they have a  strong assumption on the conditional dependence on the transition of two adjacent items. 
 
With the fast development of deep learning techniques, the neural network-based models, especially the ones that can model sequential data, have been widely applied for SBR.  Hidasi et al.\cite{SR-RNN} made an early attempt to apply RNN for SBR and designed a modal called GRU4REC, which used gated recurrent unit (GRU) to capture the transition relationship between adjacent items. Due to its gate mechanism with memory function, it can capture more complex relationships and achieve promising performance.  Later on, they proposed a parallel RNN architecture to improve the efficiency~\cite{P-RNN}. Thereafter, increasing research efforts have been devoted to developing more advanced neural network-based SBR models. Tan et al. \cite{I-RNN} reported a data enhancement method to improve the performance of RNN model; Wang et al.\cite{MCPRN} established multiple channels and used the memory unit PSRU (GRU variants) to capture different purposes in the session. In addition, Li et al.\cite{NARM} presented the NARM model, which introduces the attention mechanism into the RNN to capture the sequential behavior and main purpose of users. Liu et al.\cite{STAMP} proposed the STAMP model to combine users' general and local interests for SBR to infer the main purpose.

\begin{figure*}[th]
  \centering \includegraphics[scale=0.55]{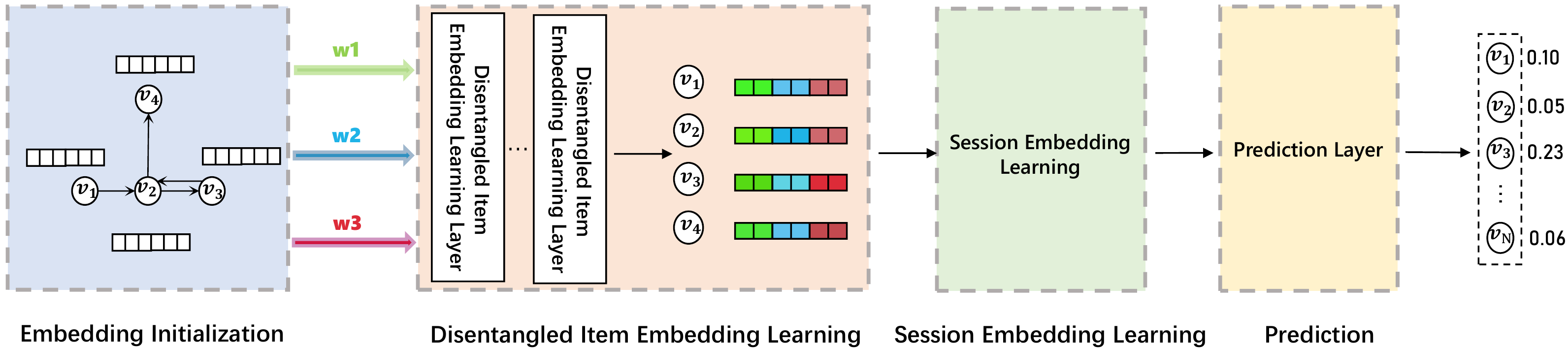}
\centering
\caption{The overview of the proposed Disen-GNN model.}
\label{allmodel}
\end{figure*}

In recent years,  graph neural networks have achieved great success in various tasks and exhibited great potential for representation learning. They have also been applied in recommendation~\cite{NGCF,Light-GCN,A2-GCN,IMPGCN,ED-GCN,BGCN}.  Wu et al.\cite{SR-GNN} applied graph neural network in SBR and proposed SR-GNN, which uses GNN with a gated mechanism to capture the complex item transition relationships by aggregating neighbor information on the graph.  Xu et al.\cite{GC-SAN} integrated the self-attention network into GNN, aiming to capture long-term dependencies with the modeling of item positions in the session. Qiu et al.\cite{FGNN} proposed an item feature encoder WGAT to set different weights to neighbors for embedding learning in the graph. Yu et al.\cite{TAGNN} extended the SR-GNN by integrating the features of the target item into the session representation learning and achieved a better performance. More recently, some researchers attempted  to consider the global session information, which utilizes the information of all sessions in the dataset to model the transition relationships between items in the target session~\cite{FGNN-BCS}\cite{GCE-GNN}. 

Although great progress has been achieved, previous methods have not considered users' multiple intents into the learning of session embeddings. As well-recognized, users' interests are diverse and the intents of a clicking behavior can be driven by different factors of the target item. Existing methods have not disentangled the multiple intents and distill the main purpose from those intents to different items in a session for representation learning. In this paper, we propose a disentangled GNN model to learn users' disentangled interests and use an attention mechanism to capture user's attention on different factors when learning the session embedding for the next item prediction.

\subsection{Disentangled Representation Learning}
Disentangled representation learning, which aims to identify and disentangle the underlying explanatory factors behind the data\cite{RL}, has gained considerable attention in various fields, such as  image and text representation learning~\cite{Infogan,Text,CPM}. In recent years, it has also been used to model user's diverse intent for recommendation \cite{ma2019learning,wang2020DisenHAN,DGCF}.  
Ma et al.~\cite{ma2019learning} applied disentangled learning to model user preference by associating difference concepts with user intentions separately.  Wang et al.~\cite{wang2020DisenHAN} proposed a disentangled heterogeneous graph attention network to learn disentangled user/item representations from different aspects over a heterogeneous information network. For studying the diversity of user intents on adopting the items, Wang et al.~\cite{DGCF} presented a GCN-based model to study users' diverse intents on adopting the items. They adopted the distance correlation to enhance the independence of different intents in embedding learning. 
In this paper, we integrate the disentangled representation learning into the gated graph neural network (GGNN) to represent items  with independent factors. By representing items with disentangled factors, we can better capture the main purpose of the user in a target session.

\section{The Proposed Model}
\subsection{Problem Statement}
Session-based recommendation aims to predict an item that the user would like to click next based on the current session. Let $\mathcal{V}=\{ v_1,v_2,\dotsc,v_N\}$  denote all the unique items involved in all sessions and $s=[v_{s,1},v_{s,2},\dotsc,v_{s,n}]$ represent an anonymous session, in which items are ordered by timestamps and $v_{s,k} \in \mathcal{V_s} (1 \leq k \leq n)$ denotes an interacted item within the session $s$. In our model, each item $v_i \in \mathcal{V}$ and each session $s$ will be embedded into the same space and let $\mathbf{e_i}\in \mathds{R}^d$ and $\mathbf{s} \in \mathbb{R}^d$ denote the representations of item $i$ and session $s$, respectively.\footnote{In the paper, we use bold uppercase letters, bold lowercase letters, and nonbold letters to denote matrices, vectors, and scalars, respectively. unless otherwise specified, all vectors are in the column form.} $d$ is the dimension of the representation vector.  For a given session $s$, the goal of session-based recommendation is to predict the next item $v_{s,n+1}$. Therefore, to recommend the desired items to the anonymous user for a given session $s$, our model aims to calculates the probabilities $\hat{y}=\{\hat{y}_1,\hat{y}_2, \dotsc, \hat{y}_N \}$ for all candidate items and select the items with the highest probabilities for recommendation. 

 For each session sequence $s$, we construct a directed session graph $\mathcal{G}_{s} =(\mathcal{V}_{s} , \mathcal{E}_{s})$, where $\mathcal{V}_{s}$ and $\mathcal{E}_{s}$ are the node set and edge set, respectively. In our setting, each node represents an item $v_{s,i}\in \mathcal{V}_s $ and a directed edge from  $v_{s,i-1}$ to $v_{s,i}$ denote that $v_{s,i}$ was visited right after $v_{s,i-1}$ in this session.
 
\subsection{Our Model}
 \textbf{Overview.} Figure~\ref{allmodel} gives an overview of our proposed disentangled graph neural network model (\textbf{Disen-GNN}) for session-based recommendation. Our model mainly consists of four components. 1) \textbf{Initialization.} In this module, each session is converted into a directed graph and each item in the session is encoded into an embedding vectors with $K$ chunks, with the assumption that each chunk represents the features of a factor.  In addition, we compute a similarity matrix for all the adjacent items based on the features of each factor. 2) \textbf{Disentangled Item Embedding Learning.} In this module, we introduce a factor-based similarity matrix, which estimates the similarity between adjacent items based on the embeddings of each factor. The similarity matrix is then integrated into the GGNN layers to learn item embedding factor-wisely. A residual attention mechanism is also designed  to keep the uniqueness of each item to prevent over-smoothing. 3) \textbf{Session Embedding Learning.}  To capture the user purpose in a session, we use an attention network to compute user's attention to different factors of each item based on the features of the last item, which represents user's local intent. With the assigned attention weights, the session embedding is generated by weightedly aggregating the factor embeddings of all items in the session.  4) \textbf{Prediction.} For each candidate item, we predict its probability by matching its embedding with the session embedding. The item with the highest probability is recommended as the next item to the user.

\subsubsection{Initialization}
\textbf{Embedding Initialization.}
 Existing session-based recommendation methods usually represent an item as a holistic representation. However, the intent of a user clicking an item is diverse~\cite{D-GCN,MCPRN}, which can be driven by different factors of the target item. An item is characterized by the features of different factors. By encoding an item with a holistic representation, it is hard to infer the particular intent of a user from different items within a session. To tackle the problem, we refer to the disentangled learning techniques to learn the item embeddings, in which the representation of each item is cast into $K$ chunks, with the assumption that each chunk represent a particular factor and those factors are independent to each other. The use of disentangled representation learning method can help us learn better and more robust item features. More importantly, the disentangled factors can facilitate the inference of user intent in a session more accurately with the use of attention mechanism. 
 
 Formally, for an arbitrary session $s=[v_{s,1},v_{s,2},\dotsc,v_{s,n} ]$, each item $v_{s,i}$ is represented by an embedding $\mathbf{e}_{s,i}\in \mathds{R}^{d}$ and the session can be represented as $\mathbf{E}_{s}=\{\mathbf{e}_{s,1},\mathbf{e}_{s,2},\dotsc,\mathbf{e}_{s,n} \}$. The embedding of each item $\mathbf{e}_{(s,i)}$ is cast into $K$ chunks and each chunk represents a latent factor:
\begin{equation}
  \mathbf{c}_{i,k} = \frac{\sigma(\mathbf{W_{k}^{\top}} \cdot \mathbf{e}_{s,i})+\mathbf{b_{k}}}{{\|\sigma(\mathbf{W_{k}^{\top}} \cdot \mathbf{e}_{s,i})+\mathbf{b_{k}}\|}_2},\label{generate disentangled representation}
\end{equation}
 where $\mathbf{W_{k} }\in\mathbb{R}^{d \times \frac{d}{K} }$ and $\mathbf{b_{k} }\in \mathbb{R}^{ \frac{d}{K} }$ are the parameters of $k ^ {th}$ factor. $\sigma$ is a nonlinear activation function. $l_2$ regularization is adopted to avoid overfitting. Accordingly, the initial embedding for the session $s$ is represented as  $\mathbf{E}_{s}^{(0)}=\{\mathbf{e}_{s,1}^{(0)},\mathbf{e}_{s,2}^{(0)},\dotsc,\mathbf{e}_{s,n}^{(0)} \}$, where $\mathbf{e}_{s,i}^{(0)} = [\mathbf{c}_{i,1}^{(0)},\mathbf{c}_{i,2}^{(0)},\dotsc,\mathbf{c}_{i,K}^{(0)}]\in \mathbb{R}^{\frac{d}{K} }$. 
 
\textbf{Factor-based Similarity Matrix.} \label{sec:simmatrix}
In the above, we represent each item by concatenating the features of different factors. Note that two items can be similar on one factor (e.g., color) but different from another factor (e.g., style). From the perspective of the transition behaviours between two adjacent items within a session (of a user), the next clicked item should be similar to the current item on the factors in which the user is interested. In light of this, we would like to define a factor-based similarity matrix to compute the similarities between the adjacent items within a session for each factor. Specifically, given a session $s$, its initial representation for the $k^{th}$ factor is denoted as $\mathbf{C}_{s,k}=[\mathbf{c}_{s,1}^{(0)},\mathbf{c}_{s,2}^{(0)},\dotsc,\mathbf{c}_{s,n}^{(0)}]$, the similarity between two adjacent items $i$ and $j$ is computed as:


 
\begin{equation} \label{eq:sim}
w_{i,j}^{k}  =\mathbf{c}_{i,k}^{\top} \cdot \mathbf{c}_{j,k}
\end{equation}
where  $\mathbf{c}_{i,k},\mathbf{c}_{j,k} \in \mathbb{R}^{\frac{d}{K}}$ represent the embeddings of $k^{th}$ factor for item $i$ and item $j$, respectively. $\mathbf{w}_{i,j}^{k}$ represents the obtained similarity scores between the adjacent item nodes $i$ and $j$. Many methods can be used to compute the similarity between two vectors. In Eq.~\ref{eq:sim},  we use the dot product to compute the similarity for simplicity. 

\begin{figure}[t]
  \centering
  \includegraphics[scale=0.5]{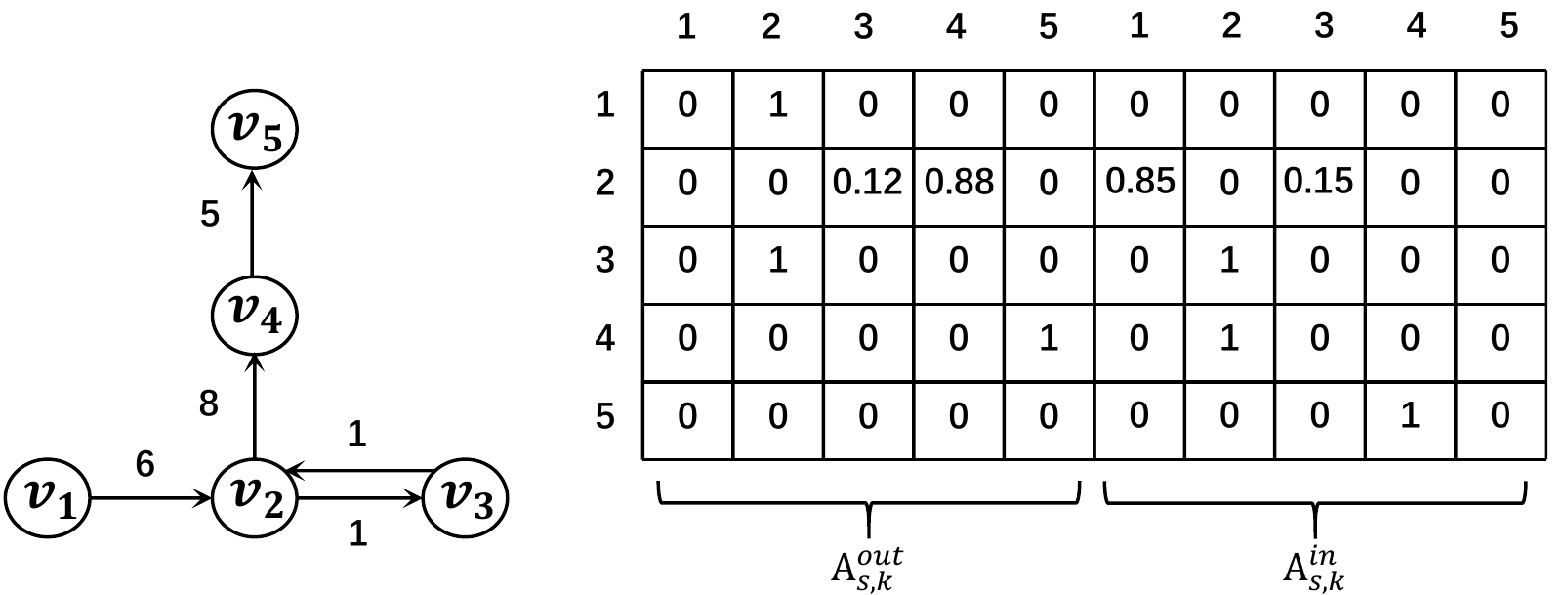}

\centering
\caption{ An example of a session graph and factor-based similarity matrix $\mathbf{A_{s,k}^{in}},\mathbf{A_{s,k}^{out}}$. The edge weight in the graph denotes similarity score, and each element in the matrices represents normalized similarity score.}
\label{matrix}
\end{figure}

Notice that the adjacent items $j \in \mathcal{N}_i$ of an item $i$ within a session $s$ are also the first-order neighboring nodes of the item node $i$ in the constructed graph $\mathcal{G}_s$ of the session $s$. And the similarity between $i$ and $j$ can be regarded as the weight for the edge between them. For the subsequent embedding learning process by using GCN techniques (as shown in Section~\ref{sec:embedding}) on $\mathcal{G}_s$, we normalize the edge weights (or similarity scores) of outgoing edges and incoming edges separately for each node. Taking the outgoing neighbors as an example, the normalized similarity score is computed as:

\begin{equation}
    \hat{w}_{i,j}^{k}  =\frac{w_{i,j}^{k}}{{\textstyle \sum_{j^{'}\in \mathcal{N}_i^o}^{\mathcal{N}_i^o}{w}_{i,j^{'}}^{k}} }
\end{equation}
where $\mathcal{N}_i^o$ denotes the set of outgoing neighbours of $i$.  After obtaining the normalized similarity scores of all adjacent items, we construct the out-degree $\mathbf{A}_{s,k}^{out}$ and in-degree similarity matrix  $\mathbf{A}_{s,k}^{in}$. Fig.{\ref{matrix}} shows a toy example for the two types of matrix.


\begin{figure*}[th]
  \centering\includegraphics[scale=0.55]{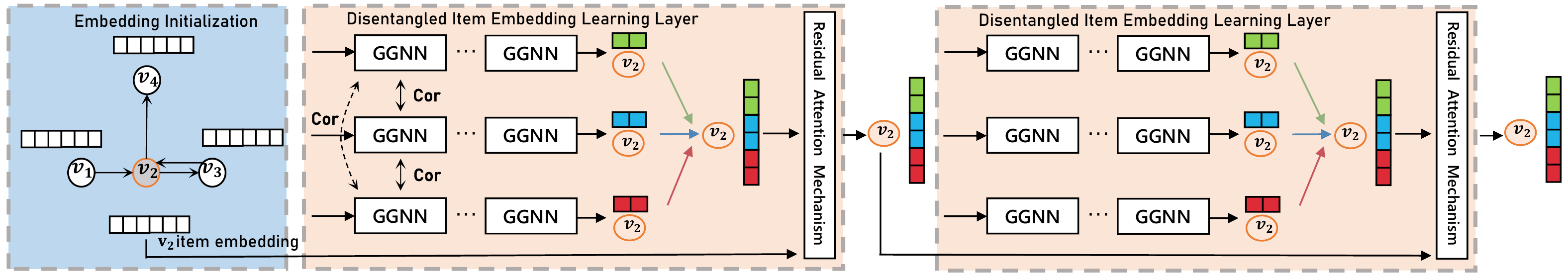}
  \centering
\caption{ An example of our disentangled item embedding learning method with two layers. In this example, all the nodes (i.e., items) in a session graph are first embedded into different subspaces (each subspace corresponds to a factor), and then learn their representations via multiple layers of GGNN. In the next stage, we take the node $v_2$ as example. The embeddings of different factors are concatenated and fused by a residual attention mechanism to obtain the item representation. Finally, the distance correlation method is used to disentangle the features of difference factors. }
\label{itememb}
\end{figure*} 

\subsubsection{ Disentangled Item Embedding Learning}
 In this section, we describe the details of our method for disentangled item embedding learning (\textbf{DIEL}). Fig.\ref{itememb} shows the structure of our method with two DIEL layers. In each DIEL layer, we firstly adopt a designed \textbf{factor-wise embedding propagation} algorithm to learn the representation of each node (i.e., item)  using multiple layers of gated graph neural networks (GGNN) for different factors separately; and then employ the \textbf{distance correlation} method to disentangle the learned features of different factors, aiming to make those factors independent to each other. The initialized embeddings of items are used as input to the first DIEL layer, and the output of this layer is used as input for the next DIEL layer We can stack multiple DIEL layers to learn better item embeddings. In each DIEL layer, we employ a \textbf{residual attention mechanism} to keep the unique features of each node during the multiple layers of embedding propagation and updating.
 
\textbf{Factor-wise Embedding Propagation.} \label{sec:embedding}
To learn the disentangled item embeddings, we employ the graph neural network techniques~\cite{GNN}\cite{GNN-R} to update the item embedding factor-by-factor. In particular, we adopt the SR-GNN~\cite{SR-GNN} because of its success. The difference is that we replace the adjacency matrix in SR-GNN \cite{SR-GNN}  with the factor-aware similarity  matrix $A_{s,k}^{in},A_{s,k}^{out}$ defined in Section~\ref{sec:simmatrix}. Let $L$ be the number of DIEL layer in our model, $T$ be the number of GGNN layers in each DIEL layer, $K$ be the  number of factors. We will use $l (1 \leq l \leq L)$, $t (1 \leq t \leq T)$, and $k (1 \leq k \leq K)$ to denote the $l$-th DIEL layer, $t$-th GGNN layer, and $k$-th factor, respectively. In the following equations, $\mathbf{a}^t_{i,k}$ denotes $\mathbf{a}^{t,l}_{i,k}$ and it is the same for other notations, including $\mathbf{z}^t_{i,k}$, $\mathbf{r}^t_{i,k}$, $\widetilde{\mathbf{c}^t_{i,k}}$, and $\mathbf{c}^t_{i,k}$. We omit the notation $l$ for simplicity as the operation is the same for all the DIEL layers.  The embedding propagation and updating rules for each factor are formulated as follows:

\begin{equation}
    \hspace{-0.40cm}
    \begin{aligned}
      \mathbf{a}^t_{i,k}  &=  Concat(\mathbf{H}_{in} \mathbf{C}_{s,k}^{t-1} (\mathbf{A}_{s,k,i:}^{in})^{\top} + \mathbf{b}_{in},\\
      \ & \  \ \ \ \ \  \ \ \ \ \ \  \ \ \ \ \ \  \mathbf{H}_{out} \mathbf{C}_{s,k}^{t-1} (\mathbf{A}_{s,k,i:}^{out })^{\top} + \mathbf{b}_{out}),\label{eq:node-representation}
    \end{aligned}
  \end{equation}
  \vspace{-0.45cm}
\begin{align}
  \mathbf{z}^t_{i,k} &= \sigma\left(\mathbf{W}_z\mathbf{a}^t_{i,k}+\mathbf{U}_z\mathbf{c}^{t-1}_{i,k}\right), \label{eq:update-gate}\\
  \mathbf{r}^t_{i,k} & = \sigma\left(\mathbf{W}_r\mathbf{a}^t_{i,k}+\mathbf{U}_r\mathbf{c}^{t-1}_{i,k}\right), \label{eq:reset-gate}\\
  \widetilde{\mathbf{c}^t_{i,k}} & = \tanh\left(\mathbf{W}_o \mathbf{a}^t_{i,k}+\mathbf{U}_o \left(\mathbf{r}^t_{i,k} \odot \mathbf{c}^{t-1}_{i,k}\right)\right), \label{eq:candidate-state}\\
  \mathbf{c}^t_{i,k}  &= \left(1-\mathbf{z}^t_{i,k} \right) \odot \mathbf{c}^{t-1}_{i,k} + \mathbf{z}^t_{i,k} \odot \widetilde{\mathbf{c}^t_{i,k}} , \label{eq:final-state}
\end{align}
where $\mathbf{A}_{s,k,i:}^{in},\mathbf{A}_{s,k,i:}^{out} \in\mathbb{R}^{{1} \times{n}}$ represents the $i$-th row of the similarity matrix of $\mathbf{A}_{s,k}^{in}$ and $\mathbf{A}_{s,k}^{out}$, respectively. $\mathbf{H}_{in},\mathbf{H}_{out} \in \mathbb{R}^{\frac{d}{K}\times \frac{d}{K} }$ denotes the weight matrix to be learned. $\mathbf{C}_{s,k}^{t-1}=[\mathbf{c}_{1,k}^{t-1},\dotsc,\mathbf{c}_{n,k}^{t-1}]$ is the representation of the $k^{th}$ factor of all the items in the session $s$ at the $(t-1)$-th layer of GGNN. $\mathbf{z}^t_{i,k} $  and $\mathbf{r}^t_{i,k}$ indicate the update and reset gate, respectively; $\sigma$ is the sigmoid activation function. $\odot$ indicates the element-wise multiplication operation. $[\mathbf{c}_{1,k}^{t},\dotsc,\mathbf{c}_{n,k}^{t}]$ is the learned embedding of all the nodes after $t$ layers of GGNN. Based on the success of  \cite{SR-GNN}, we expect that it can extract valuable information from neighbor nodes to learn the current node's embedding by using the updating and resetting gate mechanisms. 

After the factor-wise embedding propagation over all the GGNN layers, the embedding of an item $i$ can be obtained by concatenating the learned embeddings of all factors through all the GGNN layers. It is represented by  $\mathbf{e}^{l,T}_{s,i}=[\mathbf{c}^{l,T}_{i,1},\dotsc,\mathbf{c}^{l,T}_{i,K}]$, where $l$ is the $l$-th DIEL layer.

\textbf{Residual Attention Mechanism.}
 It is well-known that the GNN-based models suffer from the notorious over-smoothing problem~\cite{over-smoothing,IMPGCN}. Specifically, when stacking more layers, the node embeddings become indistinguishable. To alleviate the negative effect of over-smoothing problem, we propose a residual attention mechanism to update an item's embedding by aggregating the embedding learned from the neighbor nodes and its original embedding with assigned attention weights. In this way, an item can keep its own features by the residual updating scheme. Formally, the proposed residual attention mechanism is formulated as: 
\begin{align}
  &{\alpha^l}  =\mathbf{w}_{f} (\sigma (\mathbf{W}_{q}\mathbf{e}^{l-1,f}_{s,i}+\mathbf{W}_{p}\mathbf{e}_{s,i}^{l,T})),\\
  &\mathbf{e}_{s,i}^{l,f}  ={\alpha} \mathbf{e}^{l-1,f}_{s,i}+(1-{\alpha} )\mathbf{e}_{s,i}^{l,T},
\end{align}
where $\mathbf{e}^{l-1,f}_{s,i}$ is the final output embedding of the item $i$ in the previous DIEL layer (i.e., $(l-1)$-th layer) and $\mathbf{e}_{s,i}^{l,f}$ is the final output in this DIEL layer. $\mathbf{W}_{q},\mathbf{W}_{p}\in \mathbb{R}^{d\times d}$ are the weight matrices; $\mathbf{w}_{f}\in\mathbb{R}^{d}$ is a weigh vector and  $\sigma$ is the sigmoid activation function. $\alpha^l$ is the attention weight to control the amount of information to preserve.

\textbf{Distance Correlation.}
 As aforementioned, we would like that each chunk of an item embedding represents a latent factor. To comprehensively and concisely characterize the features of items, it is  desired that the features of different factors are independent to each other, to avoid information redundancy. Although we factor-wisely update the item embedding over the GGNN layers, there might be still redundancy among the representations of various factors. To further encourage the independence among factor-aware representations, we adopt the distance correlation~\cite{MTCD} as a regularizer in our model. We apply the distance correlation function on the initial representation in the first layer of GGNN in the first DIEL layer, which yields the best performance in our experiments.\footnote{It is worth mentioning that we try to deploy the regularization on different GGNN layers in different DIEL layers, and the one reported in this paper obtain the best performance. We do not try to use the regularization in more than one layer (either GGNN layer or DIEL layer) in experiments, because we cannot afford the required computational space.}  The distance correlation can make any two paired vectors independent. Formally, we deploy it in our model as:
\begin{equation}
  \mathcal{L}_{dec}=\sum^{K}_{k=1}\sum^{K}_{k'=k+1}dCor(\mathbf{C}_{s,k}^{(0,0)},\mathbf{C}_{s,k'}^{(0,0)}),\label{eq:disenLoss} \\
\end{equation}
where $\mathbf{C}_{s,k}^{(0,0)}=[ \mathbf{c}_{1,k}^{(0,0)},\dotsc, \mathbf{c}_{n,k}^{(0,0)}] \in \mathbb{R}^{n \times \frac{d}{K}}$ represents the initial representation of each item in the session $s$ for  the $k$-th factor in the first GGNN layer of the first DIEL layer (i.e., when $l=0$ and $t=0$). $dCor(\cdot)$ is the function of distance correlation and it is formulated as: 
\begin{equation}
  dCor(\mathbf{C}_{s,k}^{(0,0)},\mathbf{C}_{s,k'}^{(0,0)})=\frac{dCov(\mathbf{C}_{s,k}^{(0,0)},\mathbf{C}_{s,k'}^{(0,0)})}{\sqrt{dVar(\mathbf{C}_{s,k}^{(0,0)})\cdot dVar(\mathbf{C}_{s,k'}^{(0,0)})}},
\end{equation} 
\noindent where $dCov(\cdot)$ is the distance covariance between two matrices, and $dVar(\cdot)$ represents its own distance covariance. Please refer to \cite{MTCD} for more details.

\subsubsection{ Session Embedding Learning}
This section introduces the method to generate session embedding after obtaining the item embeddings. A common approach is to directly integrate the item embedding into a single embedding for the session by aggregation. In previous methods, because each item is represented as a holistic embedding, the session embedding via direct aggregation cannot well model the target user's intents on different factors. With disentangled factor-aware item embedding, the user's intents on different factors can be estimated by assigning attention weights to those factors of each item in the aggregation process. The factor-level attention modeling is expected to better capture a user's specific intents in a session.

\begin{figure}
\centering\includegraphics[scale=0.55]{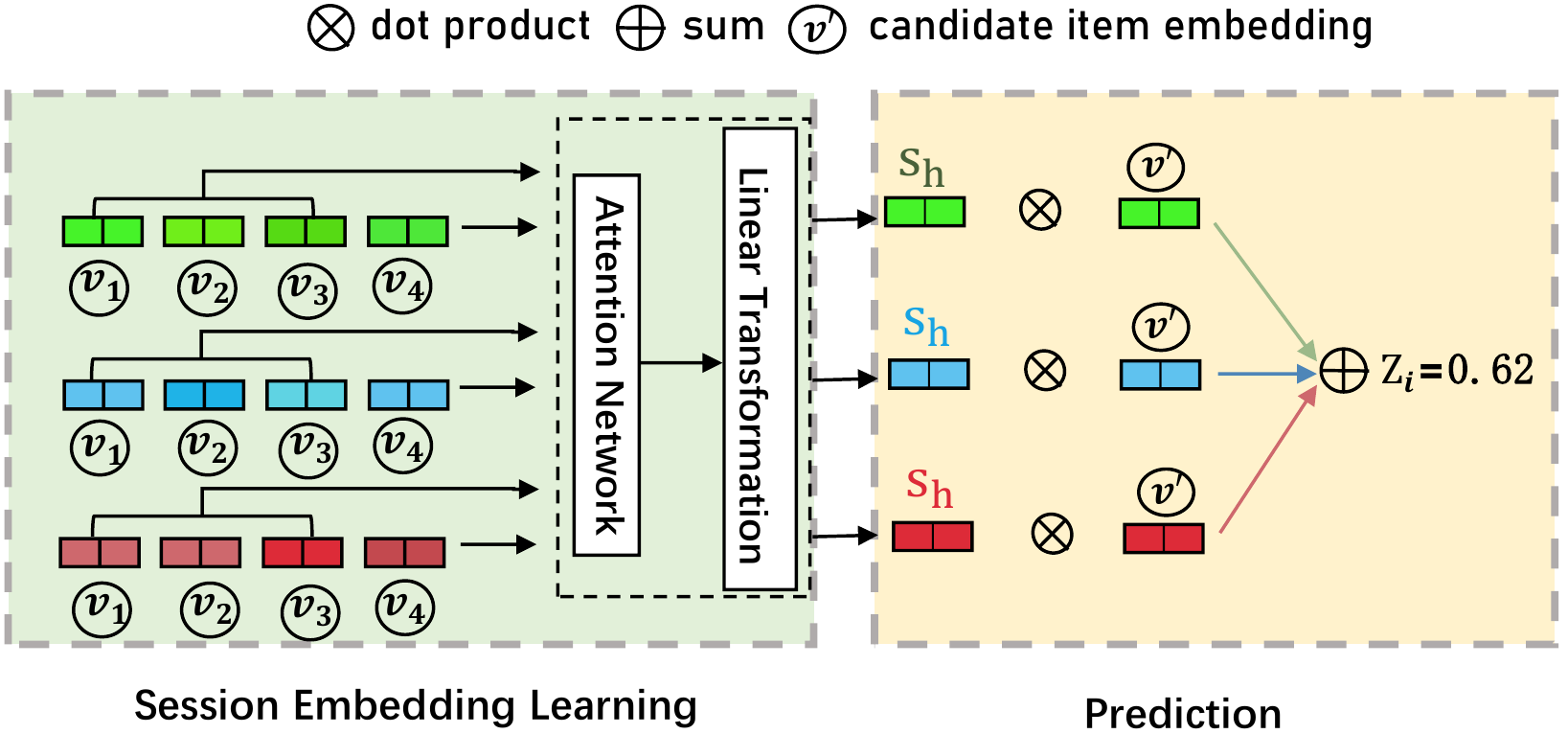}
  \centering
  \caption{Workflow of the session embedding generation and prediction.}
  \label{fig:sml}
\end{figure}

Because there could be noisy clicks  and a user's interest is dynamic along with the items clicked in a session, the combination of a user's current or local preference inferred from the last item and the global preference reflected by all the previous items can often achieve better performance~\cite{SR-GNN}. In this paper, we also adopt this strategy to generate the session embedding. Let $\mathbf{e_{s,i}}=[\mathbf{c}_{i,1},...,\mathbf{c}_{i,K}]$ be the learned item embedding for item $i$, where we omit $l$ and $t$ (i.e., $\mathbf{e_{s,i}}=\mathbf{e_{s,i}^{l,t}}$) for ease presentation. For a session $s=[v_{s,1},v_{s,2},\dotsc,v_{s,n}]$, the embedding of the last item $\mathbf{e_{s,n}}$ represents the local intent $\mathbf{s_l}$, namely, $\mathbf{s_l}= \mathbf{e_{s,n}}$. The global preference is obtained by  aggregation all the item embeddings factor-wisely with attention weights on each factor. Specifically, let  $\mathbf{s_g} = [\mathbf{s_{g,1}}, \mathbf{s_{g,2}}, \dotsc, \mathbf{s_{g,K}}]$ be the embedding of the global preference. $\mathbf{s_{g,k}}$ denotes the embedding of the $k$-th factor for $\mathbf{s_g}$ and it is computed as:
\begin{equation} \label{eq:sg}
   \mathbf{s}_{g,k}  = \sum\limits_{i = 1}^{n} {\alpha_{i,k} \mathbf{c}_{i,k}},  
\end{equation}
where $\alpha_{i,k}$ is the attention weight of the item $i$ on the $k$-th factor. An attention weight is estimated based on the embedding of the current item and that of the current preference on the $k$-th factor as:
\begin{equation} \label{eq:att}
    \alpha_{i,k} = \mathbf{q}^\top \ \sigma(\mathbf{W}_1 \mathbf{c}_{n,k} + \mathbf{W}_2 \mathbf{c}_{i,k} + \mathbf{b}),
\end{equation}
where $\mathbf{W}_1,\mathbf{W}_2\in\mathbb{R}^{\frac{d}{K} \times \frac{d}{K} }$ and $\mathbf{q}$  are the parameters to be learned in the attention neural network. From Eq.~\ref{eq:sg} and ~\ref{eq:att}, we can see that the factor-aware global preference of a session is inferred based on the factor features of each item and the current attention (with respect to the local preference) on those factors of each item. Finally, the session embedding is obtained by a linear transformation of the concatenation of the local and global preference as follow:
\begin{equation}
    \mathbf{s}_{h,k} = \mathbf{W}_3 [\mathbf{s}_{l,k}; \mathbf{s}_{g,k}],
\end{equation}
where $\mathbf{W}_3\in \mathbb{R}^{\frac{d}{K} \times \frac{2d}{K} }$ is a transformation matrix and $\mathbf{s}_h=[\mathbf{{s}_{h,1}}, \mathbf{{s}_{h,2}}, \dotsc, \mathbf{{s}_{h,K}}]$ is the final session embedding. Figure~\ref{fig:sml} shows the workflow of our session embedding generation and prediction methods. 

\subsubsection{Prediction }
 After obtaining the session representation and item embedding, for a target session $s$ and a candidate item $v_i$, the next click probability of $v_i$ is predicted by the dot product of their embeddings: 
 \begin{equation}
  \begin{aligned}
      &\mathbf{\hat{z}_i} = \mathbf{s_h}^\top \cdot \mathbf{e_{i}} = \sum^{K}_{k=1}\mathbf{s_{h,k}}^\top \cdot \mathbf{c_{i,k}},\\
      &\mathbf{\hat{y}_i} = \operatorname{softmax}\left( \mathbf{\hat{z}_i} \right),
  \end{aligned}
\end{equation}
 where $\mathbf{s}_h$ and $\mathbf{e_{i}}$  are the embeddings of session $s$ and item $v_i$, respectively. As we can see, the prediction based on dot product is actually the aggregation of the factor-level similarity between the session embedding and candidate item embedding. By further referring to Eq.~\ref{eq:sg}, it can be seen that our model considers both the factor-level similarity and the factor-level attention (i.e., $\alpha_{i,k}$) on each session item in the prediction. All the candidate items are ranked based on the descending order of the probabilities and the top ranked results are returned as recommendations.

\subsection{Loss Function }
For each session, we use the cross-entropy  as the loss function, which is formulated as:
\begin{equation}
  \mathcal{L}_{c}(\hat{\mathbf{y}}) = -\sum_{i = 1}^{m} \mathbf{y}_i \log{(\hat{\mathbf{y}_i})} + (1 - \mathbf{y}_i) \log{(1 - \hat{\mathbf{y}_i})}, 
\end{equation}
where $\mathbf{y}$ denotes the one-hot encoding vector of the ground truth value. Taking consideration of the disentangled loss, the final loss function of our model is defined as as:
\begin{equation}
  \mathcal{L}= \mathcal{L}_{c}+\lambda\mathcal{L}_{dec},
\end{equation}
where $\mathcal{L}_{dec}$ is the independence loss defined in Eq.~\ref{eq:disenLoss} and $\lambda$ is a hyper-parameter to control the regularization loss of the disentangled learning (i.e., distance correlation loss). 

\section{ Experiments }
To evaluate the effectiveness of our proposed Disen-GNN model in session-based recommendation, we perform extensive experiments on three publicly accessible datasets. Through the experiments, we mainly answer the following research questions.
\begin{itemize}
  \item{\textbf{RQ1:}} Can the proposed Disen-GNN outperform state-of-the-art session-based recommendation methods?
  \item{\textbf{RQ2:}} Does the disentangled representation positively affect our proposed model on session-based recommendation ?
  \item{\textbf{RQ3:}} How do different components of our model, such as the factor-aware similarity matrix and the gated attention mechanism, affect the performance?
  \item{\textbf{RQ4:}} How does the key parameters impact the performance of the Disen-GNN, including the number of factors and the regularization coefficient?
\end{itemize}
 
 In the next, we first introduce the experimental setup, then report and analyze the experimental results to answer the above questions sequentially.
 
\subsection{ Experimental Setup }
\textbf{Datesets.} Three widely used real-world datasets are adopted for evaluation in our experiments, including Diginetica\footnote{\url{http://cikm2016.cs.iupui.edu/cikm-cup}}, Yoochoose\footnote{\url{http://2015.recsyschallenge.com/challege.html}} and Nowplaying\footnote{\url{http://dbis-nowplaying.uibk.ac.at/#nowplaying}}. 1) The Diginetica dataset records typical transaction data of users and comes from the CIKM Cup 2016. 2) Yoochoose dataset contains user clicking behaviors on the e-commerce website of Yoochoose.com, which is released by the RecSys Challenge 2015. 3) Nowplaying is released by \cite{NOW} and contains users' listening behaviors extracted from Twitter.

We follow the commonly adopted procedures as \cite {SR-GNN}\cite {GCE-GNN} to process the datasets. Specifically, we remove the sessions with only one item and infrequent items which appears less than 5 times in each dataset. Similar to previous works~\cite{GCE-GNN,STAMP},  the sessions of the last day are used as the test data for Yoochoose and the sessions of the last week are used as the test data for both Diginetiva and Nowplaying datasets. Due to the large size of Yoochoose, only the most recent 1/64 data is used in experiments. For the other two datasets, the remaining data is used as the training data. In addition, for a session $\textit{S=}[v_{s,1},v_{s,2},...,v_{s,n}]$, we generate a set of seuqences and corresponding labels  by a splitting preprocessing, i. e. $([v_{s,1}],v_{s,2}),([v_{s,1},v_{s,2}],v_{s,3}),...,([v_{s,1},...,v_{s,n-1}],v_{s,n})$, in which $[v_{s,1},...,v_{s,n-1}]$ denotes a generated sequence and $v_{s,n}$ is the next-clicked item (i.e., the label). After the processing steps, the statistics of the three datasets used for experiments are shown in Table~\ref{tab:dataset-statistics}.
\begin{table}
	\centering
	\caption{Statistical results of datasets}
	\resizebox{\columnwidth}{!}{
	\begin{tabular}{c c c c}
    \hline	
    \toprule
		Statistics & {\it Diginetica} & {\it Yoochoose 1/64} & {\it Nowplaying} \\ \midrule
		 \# clicks & 982,961 & 557,248 & 1,367,963  \\
		\# training sessions & 719,470 & 369,859 & 825,304  \\
		\# test sessions & 60,858 & 55,898 & 89,824 \\
		\# items & 43,097 & 16,766 & 60,417  \\
		Avg. length & 5.12 & 6.16  & 7.42   \\
		Max length & 70 & 146 & 30 \\
		\bottomrule
    \hline
	\end{tabular}
	}
	\label{tab:dataset-statistics}
\end{table}

\textbf{Evaluation Metrics.} Two widely used evaluation metrics in session-based recommendation are adopted in our evaluation:\textbf{P@20} (Precision) and \textbf{MRR@20} (Mean Reciprocal Rank). P@20 measures the recommendation accuracy and MRR@20 considers the ranking quality. Specifically, \emph{P@20 denotes the percentage} and \emph{MRR@20 is the average of the reciprocal ranks} of the correctly recommended items in the top 20 results, respectively..

\textbf{Baseline methods.} We compare the proposed Disen-GNN model to a set of representative and state-of-the-art session-based recommendation methods to evaluate its effectiveness. We briefly introduce those methods as follows.
\begin{itemize}[leftmargin=*]
  \item \textbf{POP} is the popularity-based method, which recommends items simply based on their popularity in the dataset.
  \item \textbf{Item-KNN}\cite{ITEM-KNN} uses the similarity between items in the session to make recommendations. We use cosine similarity in our implementation.
  \item \textbf{FPMC}\cite{FPMC} is a sequential recommendation method based on Markov chain. Following the previous work,  the user latent representation is ignored when computing recommendation scores. 
  \item \textbf{GRU4REC}\cite{SR-RNN} is a RNN-based model by employing GRU  units to capture sequential transition between items.
  \item \textbf{NARM}\cite{NARM} introduces the attention mechanism into RNN to capture user purpose in a session. The main purpose is combined with the sequence behavior features to generate the final representation for next item prediction.
  \item \textbf{STAMP}\cite{STAMP} employs an attention-based MLP model to combine user's general interest and current interest of the last item in the session for recommendation.
  \item \textbf{SR-GNN}\cite{SR-GNN} is a state-of-the-art GNN method for session-based recommendation. Similar to STAMP, it also adopts an attention mechanism to combine user's general interest and current interest to predict the next item.
  \item \textbf{TAGNN}\cite{TAGNN} is a variant of SR-GNN\cite{SR-GNN}. It uses a target-aware attentive network to generate the session embedding. Specifically, when predicting the probability of clicking on a candidate item, the features of this candidate item is considered into the generation of the session representation.
\end{itemize}

\textbf{Parameter settings.} For the compared methods, we follow the guidelines described in the corresponding papers to preprocess the data and set the parameters. We tune the parameters carefully and report the best performance. We train the proposed Disen-GNN model by using the Back-Propagation Through Time (BPTT) algorithm \cite{BPTT}. The dropout strategy is applied between disentangled item embedding learning layers to prevent overfitting. In the training of our model, the batch size is set to 100. We employ the Adam algorithm~\cite{Adam} for optimization and set the initial learning rate to 0.005 with a decay rate of 0.1 for every 3 epochs. All parameters are initialized using a Gaussian distribution with a mean of 0 and a standard deviation of 0.1. Besides, The dropout ratio  and the weight of L2 regularization are set to 0.1 and  $10^{-5}$, respectively.  The above parameter settings are the same for all the three datasets. Other key parameters which have a large impact on the performance are tuned for each dataset, such as the number of factors (\# factors), the number of GGNN layers in a disentangled item embedding learning (DIEL) layer, etc. Details setting of those parameters are shown in Table~\ref{tab:Parameter}. All the parameters are tuned on a validation dataset, which is a random 10\% subset of the training set. 

\begin{table}
	\centering
	\caption{Parameter settings for datasets}
	\begin{threeparttable}
	\resizebox{\columnwidth}{!}{
	\begin{tabular}{c c c c}
    \hline	
    \toprule
		Statistics & {\it Diginetica} & {\it Yoochoose 1/64} & {\it Nowplaying} \\ \midrule
		 embedding size ($d$) & 80 & 100 & 96  \\
		 \# factors ($K$) & 5 & 5 & 12  \\
		 \# GGNN layers ($T$) & 2 & 3 & 3  \\
         \# DIEL layers ($L$)& 2 & 2  & 2   \\
     $\lambda$ & 5 & 10  & 15   \\
		\bottomrule
    \hline
	\end{tabular}
	}
	\end{threeparttable}
	\label{tab:Parameter}
\end{table}

\subsection{ Performance Comparison (RQ1)}
Table {\ref{tab:result-baseline-algorithms}} reports the performance comparison between our model and baselines on three datasets in terms of P@20 and MRR@20. We highlight the best and second best in bold and underlined form, respectively. It can be seen that our Disen-GNN model achieves the best performance across all three datasets in terms of the two metrics consistently. In particular, Disen-GNN obviously outperforms all the baselines by a large margin on the Diginetica and Nowplaying datasets. The results demonstrate the superiority of our model.

For the three traditional methods (i.e., POP, Item-KNN, and FPMC), the performance of POP is the worst, which is unsurprising, as it recommends items based on their popularity in the datasets without considering users' purpose in each session.  FPMC combines the first-order Markov chains and matrix factorization. Because it ignores user latent representation in session-based recommendation, its performance is also limited. Item-KNN achieves the best performance amongst the traditional methods. It recommends items based on item similarity without considering the sequential relations between items, and thus it cannot model the transitions between items.

\begin{table}
	\centering
	\caption{Performance comparisons between our model and the competitors  over three datasets. Bold and underlined text indicate best and second-best results, respectively.}
	\resizebox{\columnwidth}{!}{
	\begin{tabular}{c|c|c|c|c|c|c}
    \thickhline
	{\multirow{2}{*}{Model}} & \multicolumn{2}{c|}{Diginetica} & \multicolumn{2}{c|}{Yoochoose 1/64} & \multicolumn{2}{c}{Nowplaying} \\\cline{2-7} 
		& P@20 & MRR@20 & P@20 & MRR@20 & P@20 & MRR@20 \\ \thickhline
	  POP & 0.89  & 0.20  & 6.71  & 1.65  & 2.28  &  0.86 \\
    Item-KNN & 35.75 & 11.57 & 51.60 & 21.81  &  15.94 & 4.91 \\
    FPMC & 26.53 & 6.95 & 45.62 & 15.01 & 7.36 &  2.82 \\
    GRU4REC & 29.45 & 8.33 & 60.64 & 22.89 & 7.92 &  4.48 \\
    NARM & 49.70 & 16.17 & 68.32 & 28.63 & 18.59 & 6.93 \\
    STAMP & 45.64 & 14.32 & 68.74 & 29.67  &  17.66 &  6.88 \\
    SR-GNN & 51.26 & 17.66&  70.57 & 30.94 &  17.76 &  7.49 \\
    TAGNN & \underline{51.53} &  \underline{17.90} &  \underline{71.02} & \underline{31.12} &  \underline{19.02} &  \underline{7.82} \\ 
    Disen-GNN & {\bf 53.79} &  {\bf18.99} &  {\bf71.46} & {\bf 31.36} &  {\bf 22.22} & {\bf8.22} \\
    \thickhline
	\end{tabular}
	}
	\label{tab:result-baseline-algorithms}
\end{table}

The performance of neural network based models often performs better than the above traditional methods. GRU4REC is the first RNN-based method to model the sequential patterns for session-based recommendation. It achieves better performance over FPMC, however, it underperforms  Item-KNN on Diginetica and Nowplaying datasets.  This is because RNN is designed for sequential modeling, and session-based recommendation involves the inference of user preference in each session. Note that a user's main purpose is different for different sessions and could even be changed within a session. The subsequent NARM model introduces the attention mechanism to combine user's general interest and local interest to infer user's purpose in the session, achieving substantial improvement over GRU4REC.  STAMP adopts a simple MLP structure with a complete attention-based method and incorporates a self-attention on the last item within each session to model the local interest. The comparable performance between NARM and STAMP demonstrates the importance of differentiating the contributions of different items for session embedding generation. 

Compared with the neural network-based models, the recently developed GNN-based models further improve the performance. By iteratively aggregating feature information from local graph neighbors, GNN-based models can distill additional information from high-order neighbors over the graph structure to learn better item representations. SR-GNN applies the GNN to session-based recommendation and also adopts the self-attention method on the last item to generate session embedding. Attributed to the advanced representation learning techniques, SR-GNN outperforms previous models.  TAGNN further improves the SR-GNN by considering the features of the target item in session embedding generation and can achieve better performance over SR-GNN in general (besides P@20 on the Nowplaying dataset). The good performance of SR-GNN and TAGNN demonstrates the advantage of applying GNN in session-based recommendation for embedding learning. 

 Inspired by the effectiveness of the GNN models, our proposed model also adopts the GNN technique to learn item embeddings. In particular, we disentangle the item embeddings into independent factors and separately update the embedding of each factor via propagation over the session graph. In addition, the last item (which represents local interests) of the session is used to estimate the importance of each factor of all the items to generate the session embedding. With those specially designed components, our model outperforms the best compared method (i.e., TAGNN), especially on the Diginetica and Nowplaying datasets. Concretely, our model achieves a relatively improvement of 5\% and 13\% over TAGNN on the Diginetica and Nowplaying datasets, respectively. Note that the max length of sessions in the Yoochoose is much longer (see Table~\ref{tab:dataset-statistics}), and thus it requires more memory in modeling.  Due to our limited computational resources, we can only try at most five factors in Disen-GNN on this dataset, which may not be the best factor number. In the next, we will investigate the contributions of different components to the performance of our model. 
 

\begin{table}[t]
	\centering
  \caption{The effectiveness of disentangled representation learning on session-based recommendation.}
  \begin{threeparttable}
  \scalebox{1}{
  \resizebox{\columnwidth}{!}{
  \begin{tabular}{c|c|c|c|c|c|c}
    \thickhline
	{\multirow{2}{*}{Model}} & \multicolumn{2}{c|}{Diginetica} & \multicolumn{2}{c|}{Yoochoose 1/64} & \multicolumn{2}{c}{Nowplaying}   \\\cline{2-7} 
		   & P@20  & MRR@20 & P@20  & MRR@20 & P@20 & MRR@20    \\ \thickhline
    SR-GNN  & 51.26  &17.66  &70.57  & 30.94 &17.76 & 7.49\\
    TAGNN   & 51.53  &17.90  &71.02  & 31.12 & 19.02 &7.82 \\ 
    Disen-SRGNN & 51.52 & 17.56& 70.65 & 31.30& 18.76 &7.78     \\
    Disen-TAGNN & 53.26 & 18.56 &  -    & -    & 19.53 & 7.99 \\
    Disen-GNN & {\bf 53.79} &{\bf 18.99}& {\bf 71.46}&{\bf 31.36}& {\bf 22.22} & {\bf 8.22} \\
    \thickhline
	\end{tabular}
	}}
  \begin{tablenotes}
    \scriptsize
    \item[1] We cannot run Disen-TAGNN on Yoochoose due to insufficient memory.
  \end{tablenotes}
\end{threeparttable}
\label{tab:disenexp}
\end{table}

\subsection{Effectiveness of disentangled representation (RQ2)}
The main contribution of our model is to introduce the disentangled representation learning (DRL) into session-based recommendation to learn item embedding, and thus to model user interest in a session by assigning factor-level attentions to items for session embedding learning. To study the effectiveness of the DRL on session-based recommendation, we apply  DRL to two existing methods SR-GNN and TAGNN, which are also the best baselines in our experiments. 

Let Disen-SRGNN and Disen-TAGNN denote the models which apply DRL into the item embedding learning (i.e., casting item embedding into $K$ factor embedding and adding the distance correlation loss function) in SR-GNN and TAGNN, respectively. Table~\ref{tab:disenexp} shows the experimental results. Apparently, Disen-SRGNN and Disen-TAGNN outperform their counterparts consistently across different datasets, which is purely credited to the capability of disentangled representation learning on generating more comprehensive and robust item embedding.  The results can provide strong evidence to validate the potential of applying disentangled learning technique in session-based recommendation. The better performance of Disen-GNN over Disen-SRGNN and Disen-TAGNN further demonstrates the superiority of our model. The better performance is attributed to our special designs of factor-aware item embedding learning (such as the use of factor-wise similarity matrix in GGNN and residual attention mechanism). 




\begin{table}[t]
	\centering
	\caption{Ablation study of the effectiveness of factor-based similarity matrix.}
	\scalebox{1}{
  \resizebox{\columnwidth}{!}{
	\begin{tabular}{c|c|c|c|c|c|c}
    \thickhline
	{\multirow{2}{*}{Model}} & \multicolumn{2}{c|}{Diginetica} & \multicolumn{2}{c|}{Yoochoose 1/64} & \multicolumn{2}{c}{Nowplaying}   \\\cline{2-7} 
                  & P@20 & MRR@20 & P@20 & MRR@20 & P@20 & MRR@20  \\ \thickhline
    Disen-GNN-io & 53.52  &18.98  & 71.29  &{\bf 31.42} & 21.82  &7.89\\
    Disen-GNN & {\bf 53.79}  & {\bf 18.99} &{\bf 71.46}  &  31.36 &{\bf 22.22}  & {\bf 8.22} \\ 
    \thickhline
	\end{tabular}
	}}
	\label{tab:Similarity Adjacency Matrix}
\end{table}

\begin{figure}[t]
  \subfloat[Diginetica]{
	\begin{minipage}{0.49\linewidth}
		\centering
		\includegraphics[width=0.99\linewidth]{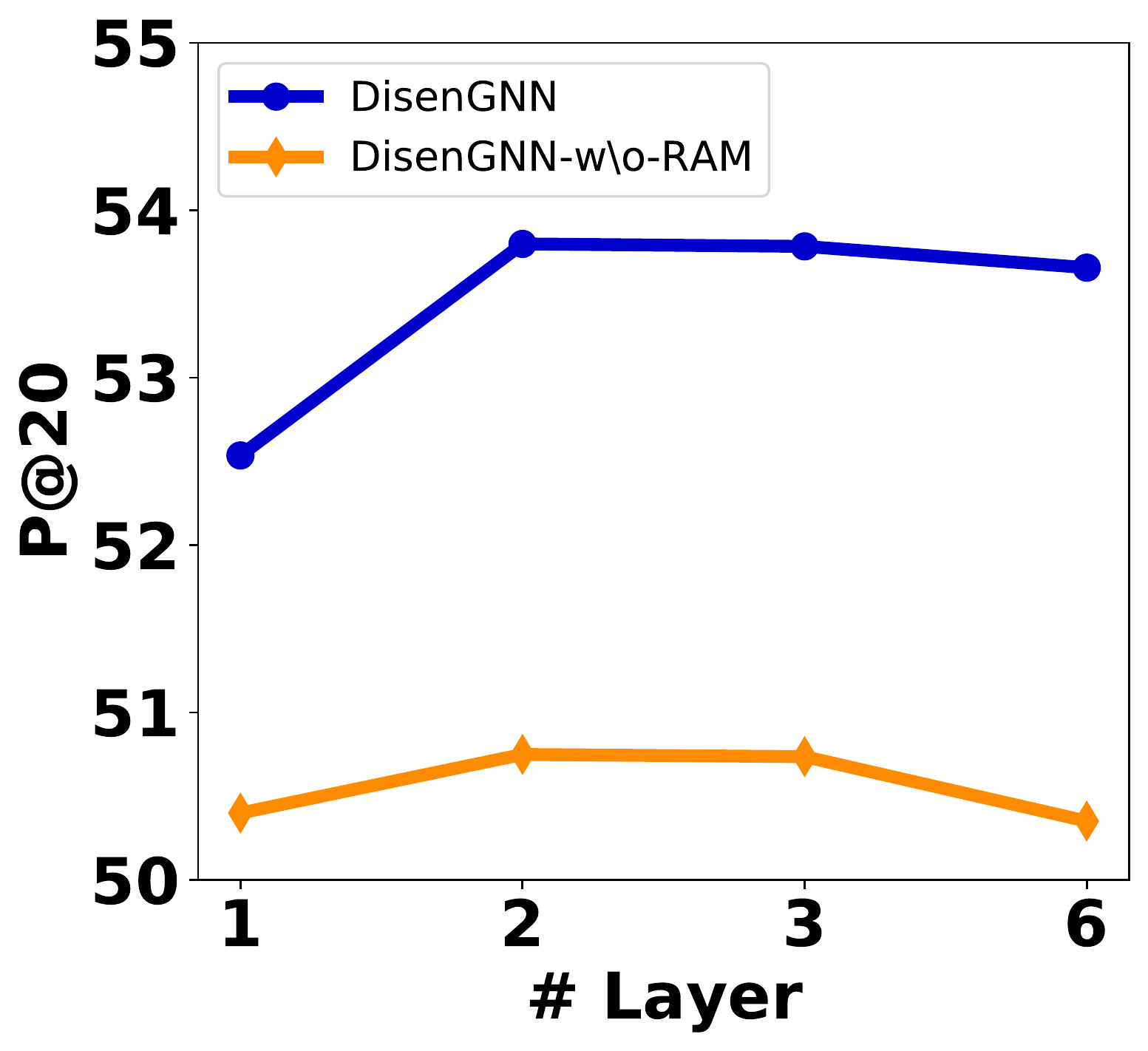}
	\end{minipage}
	\begin{minipage}{0.51\linewidth}
		\centering
		\includegraphics[width=0.99\linewidth]{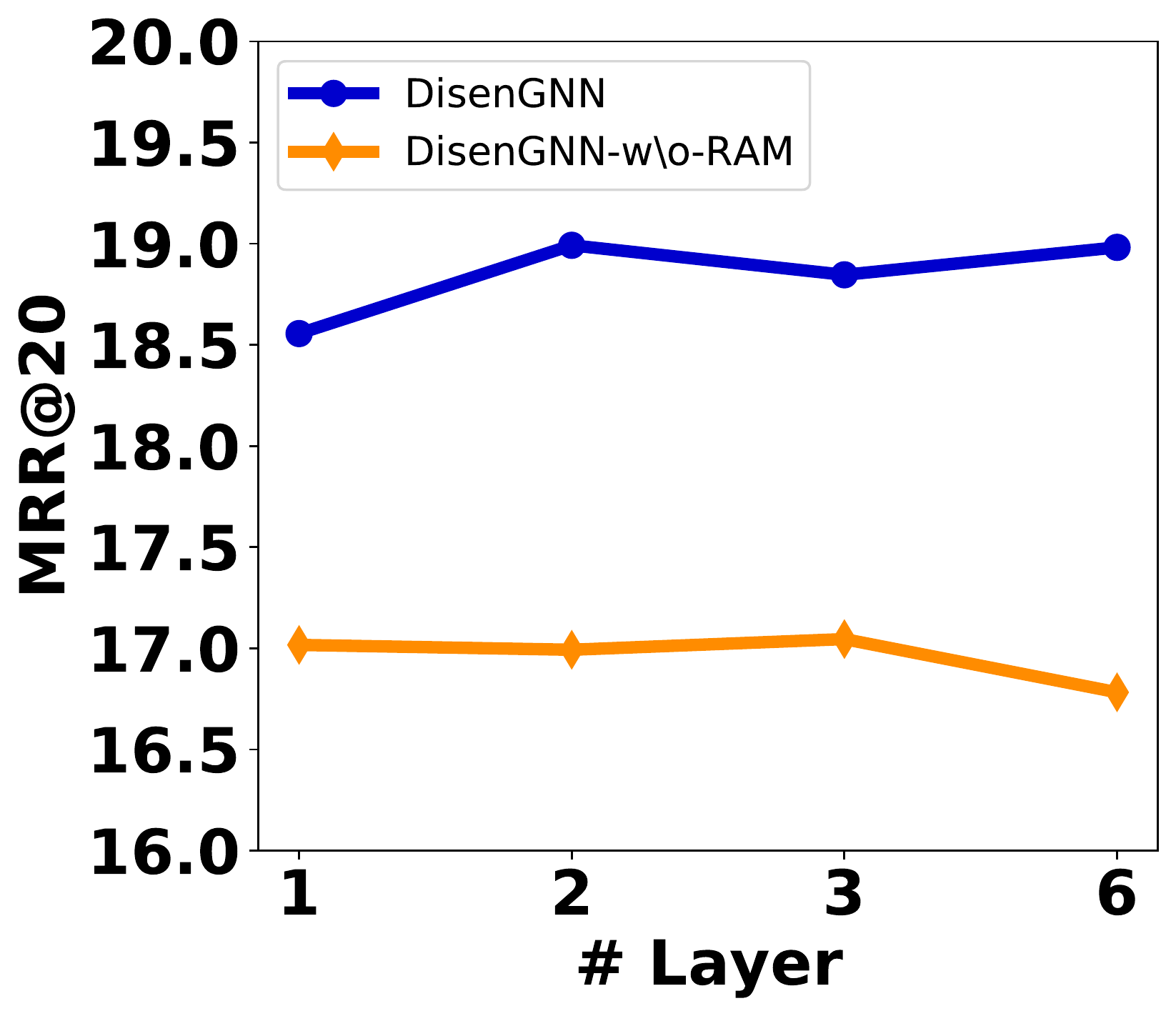}
	\end{minipage}
  }\\
	\subfloat[Nowplaying]{
  \begin{minipage}{0.49\linewidth}
		\centering
		\includegraphics[width=0.99\linewidth]{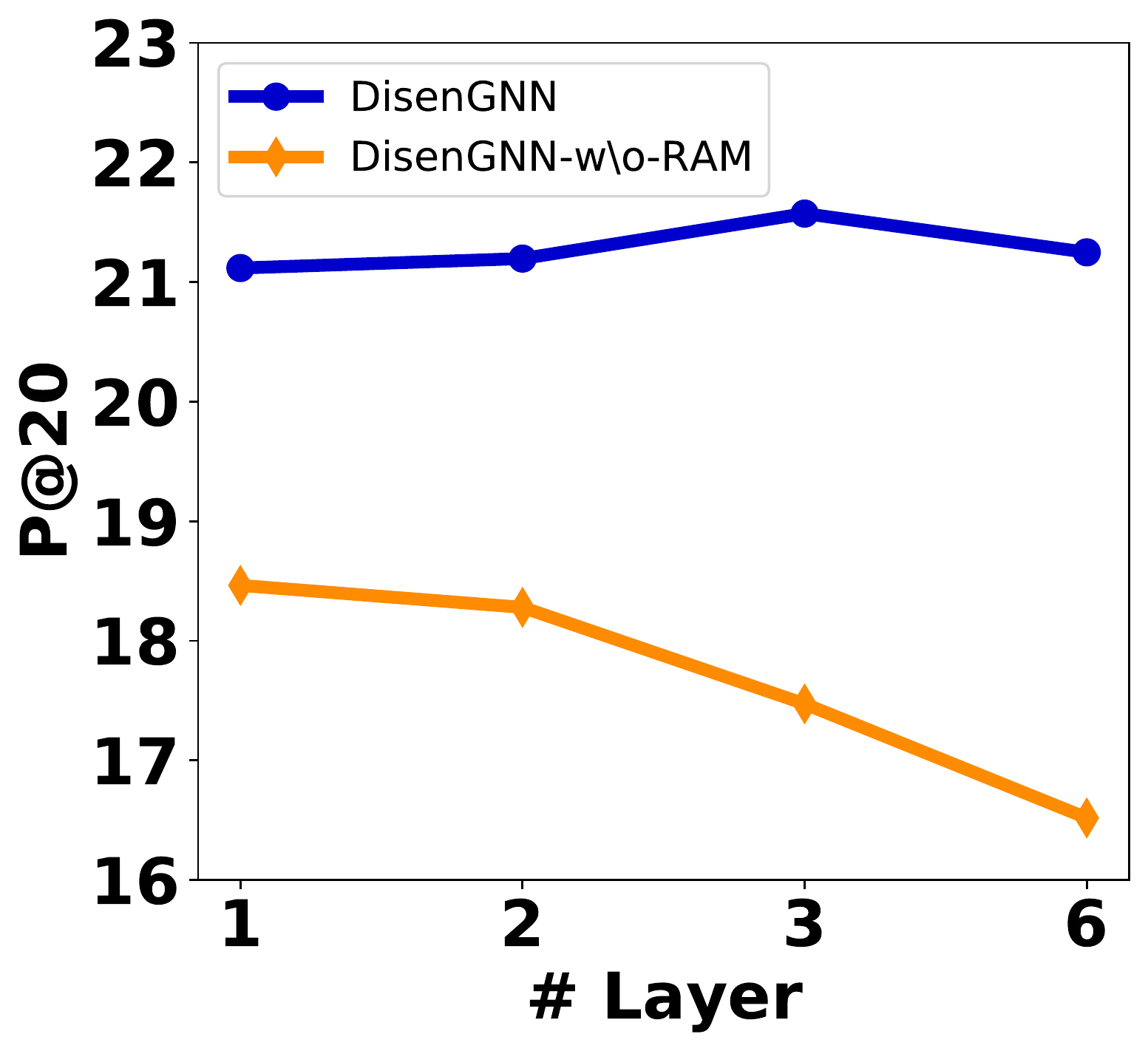}
	\end{minipage}
	\begin{minipage}{0.51\linewidth}
		\centering
		\includegraphics[width=0.99\linewidth]{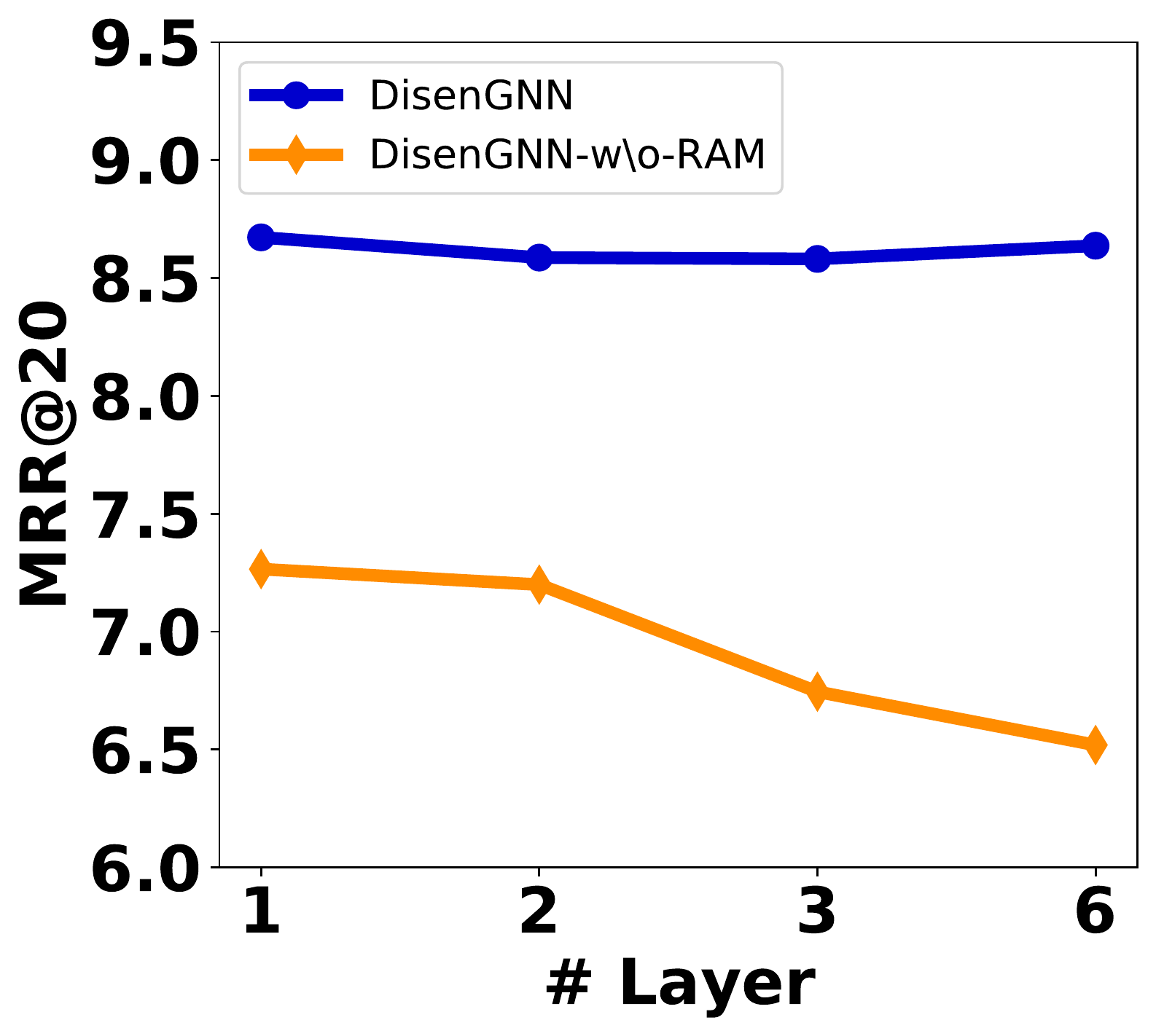}
	\end{minipage}
  }
  \caption{Effectiveness of the residual attention mechanism.}
  \label{tab:ram}
  \vspace{-10pt}
\end{figure}

\subsection{Ablation Study (RQ3)}

\textbf{Effectiveness of Factor-based Similarity Matrix. }
In the GGNN model for learning item embedding, we replace the original adjacency matrix by our defined factor-based similarity matrix (denoted by FSM, see section~\ref{sec:embedding} for details) to learn the embeddings for each factor of an item. To examine its effectiveness, we compare our model with FSM and the one with the original adjacency matrix~\cite{SR-GNN}. Table \ref{tab:Similarity Adjacency Matrix}  shows the results of Disen-GNN and the variant (i.e., Disen-GNN w/o fsm) on three datasets. The better performance of Disen-GNN indicates that using the factor-based similarity matrix to consider the importance of neighbors on each factor can facilitate learn better item embeddings and thus improve the final recommendation performance.

\begin{figure*}[ht]
	\centering
	\subfloat[P@20 on Diginetica]{\includegraphics[width=0.25\linewidth]{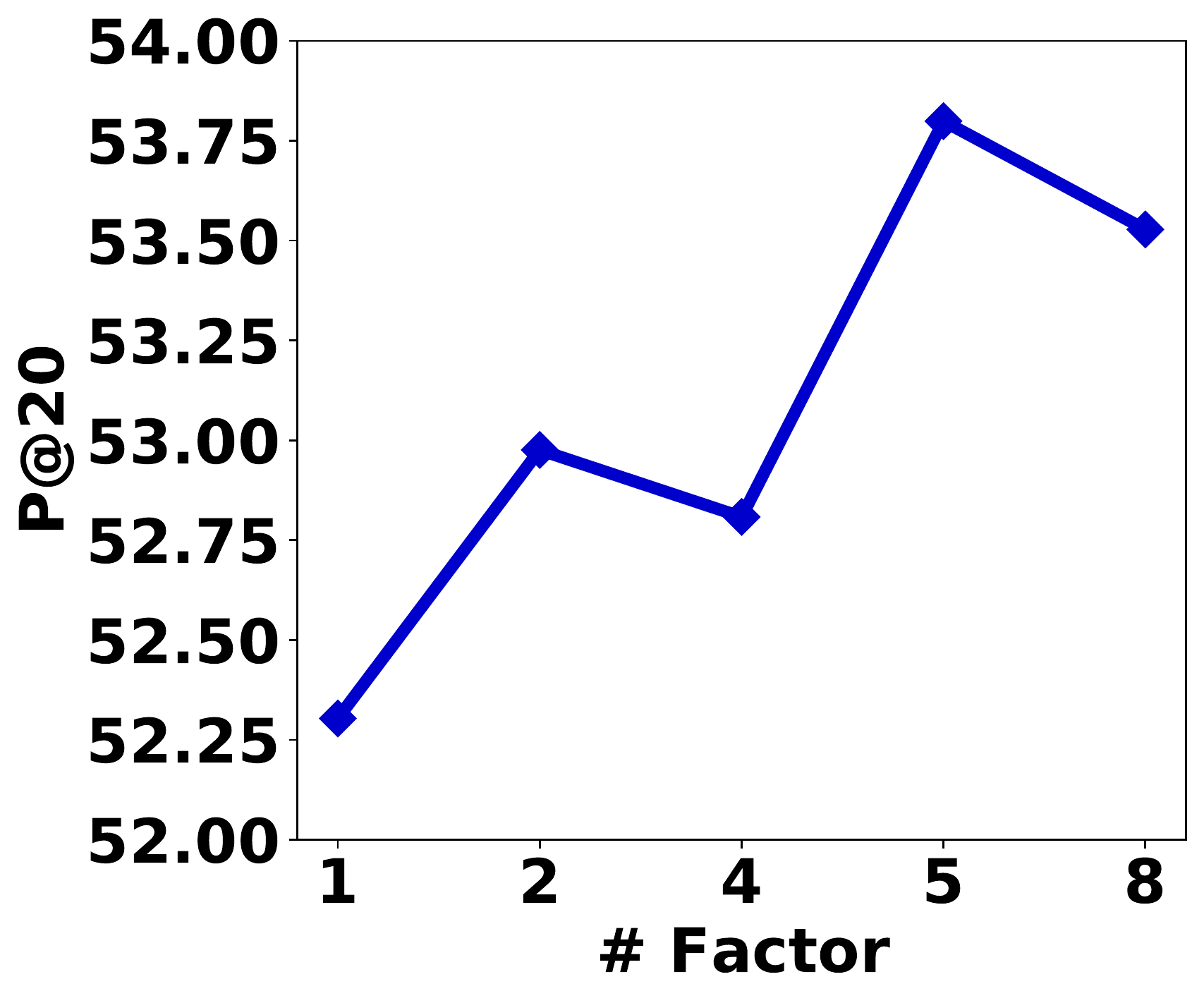}}
	\subfloat[MRR@20 on Diginetiva]{\includegraphics[width=0.25\linewidth]{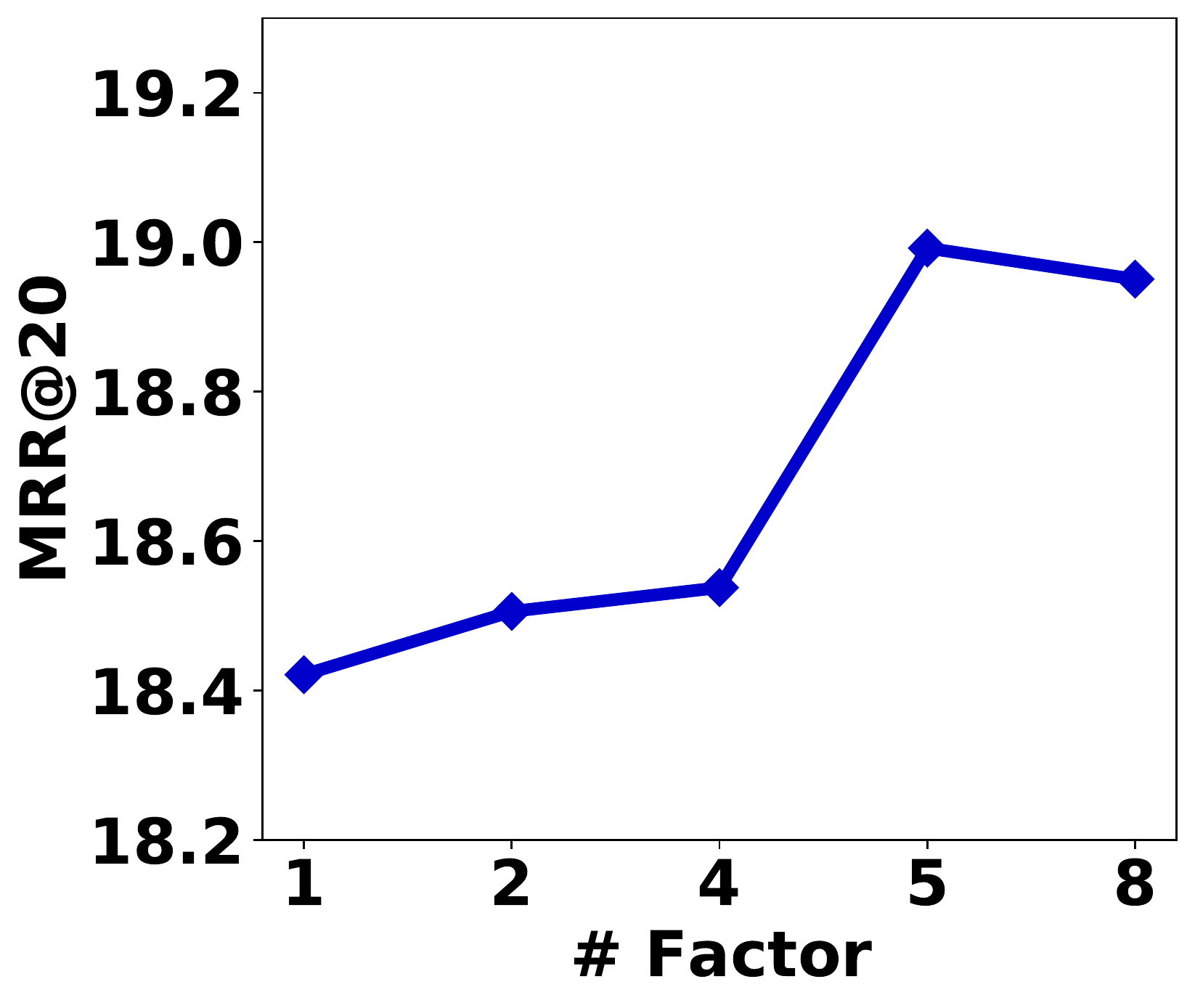}}
	\subfloat[P@20 on Nowplaying]{\includegraphics[width=0.25\linewidth]{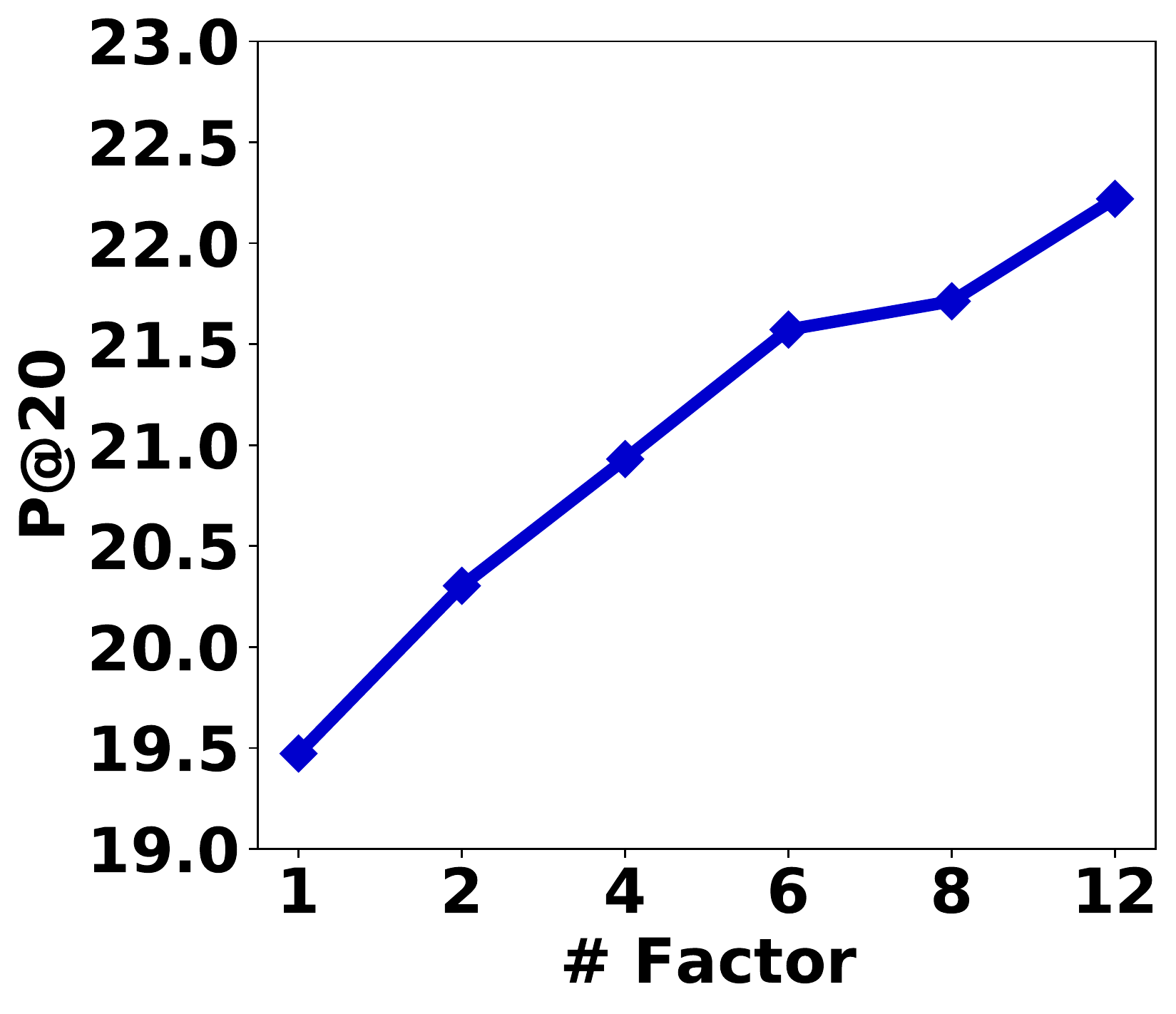}}
	\subfloat[MRR@20 on Nowplaying]{\includegraphics[width=0.25\linewidth]{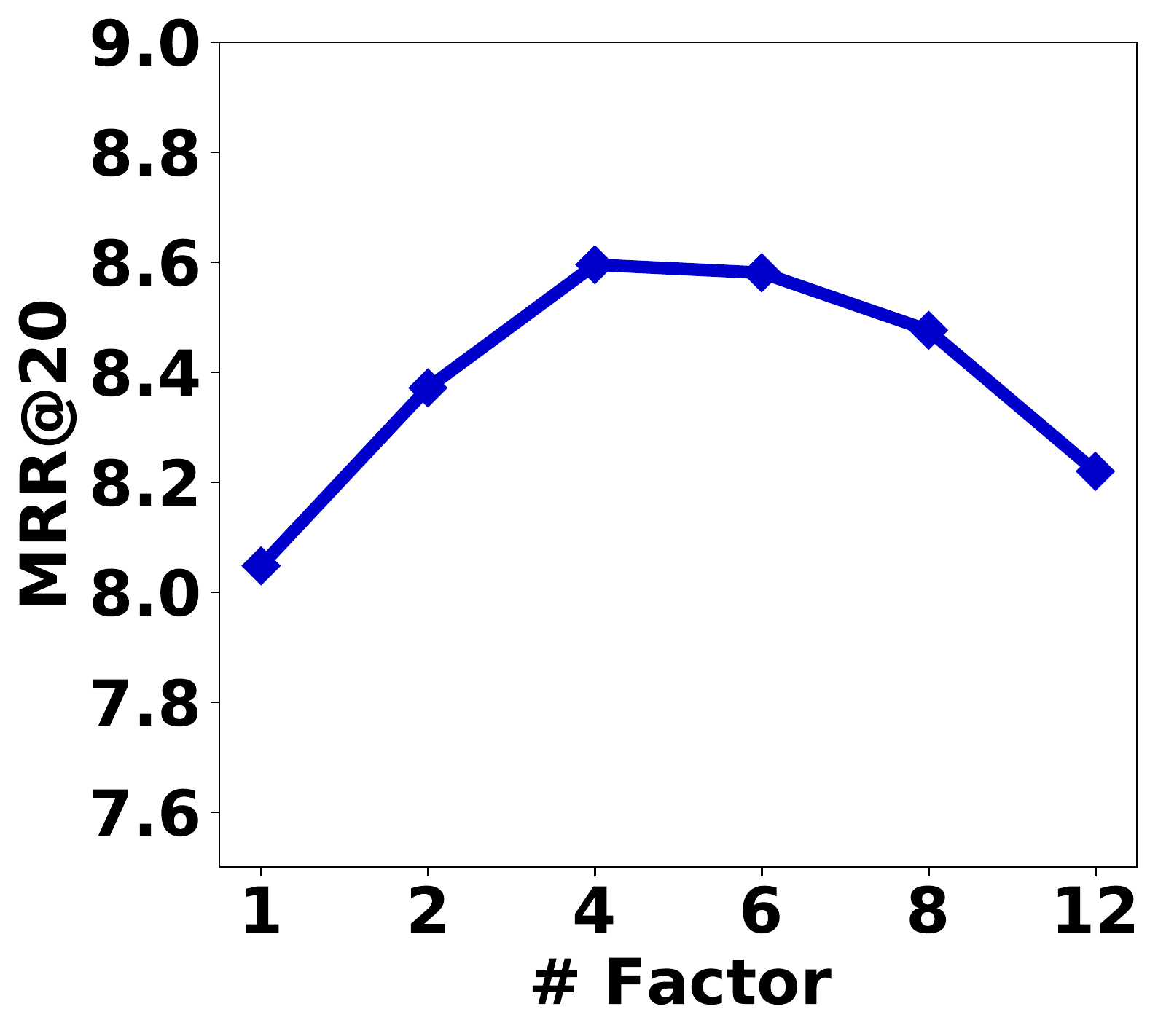}}
	\vspace{-6pt}
	\caption{Impact of factor number ($K$).}
	\vspace{-12pt}
	\label{fig:feature}
\end{figure*}

\begin{figure}[t]
	\centering
	\subfloat[P@20 on Diginetica]{\includegraphics[width=0.5\linewidth]{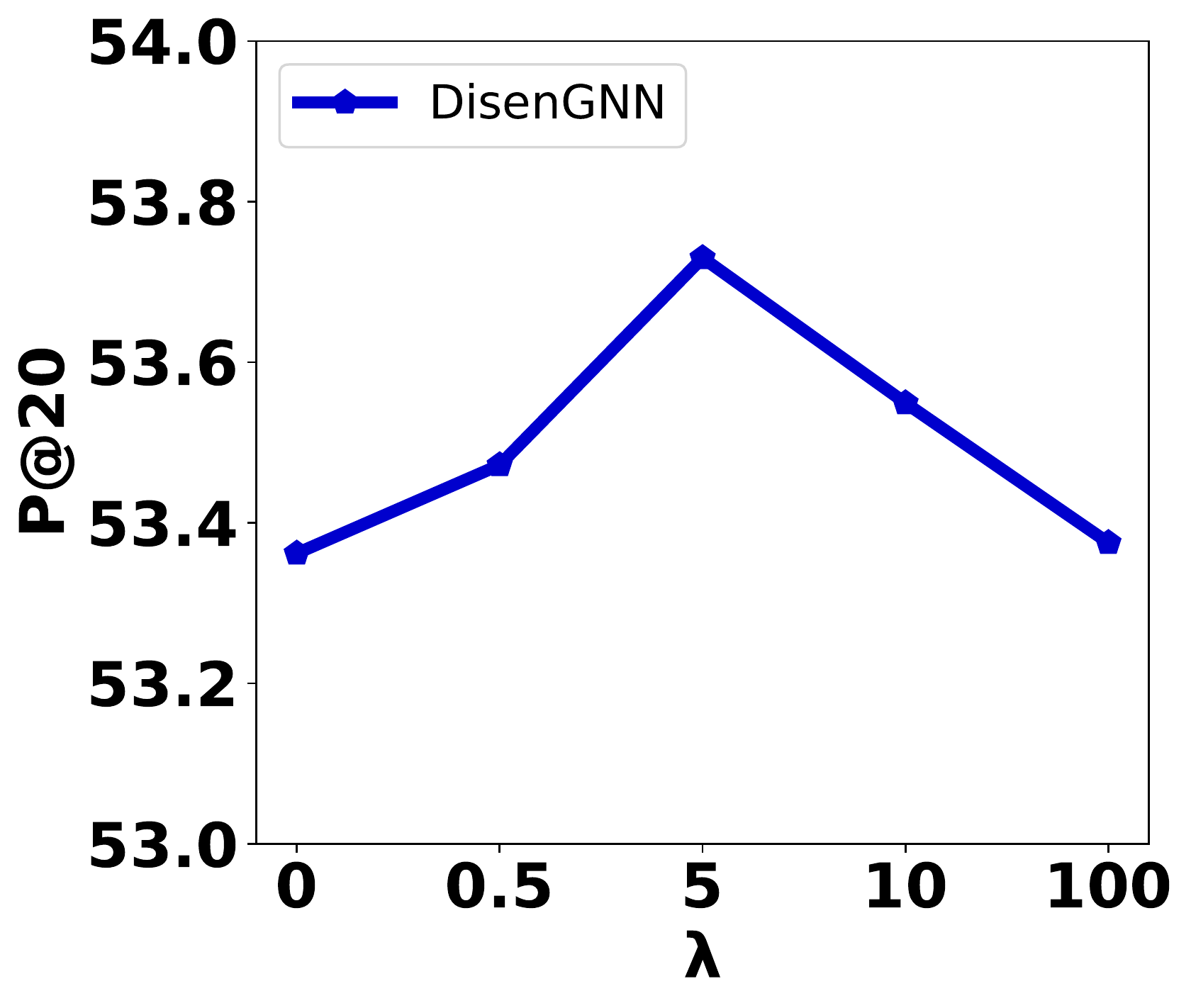}}
	\subfloat[P@20 on Nowplaying]{\includegraphics[width=0.5\linewidth]{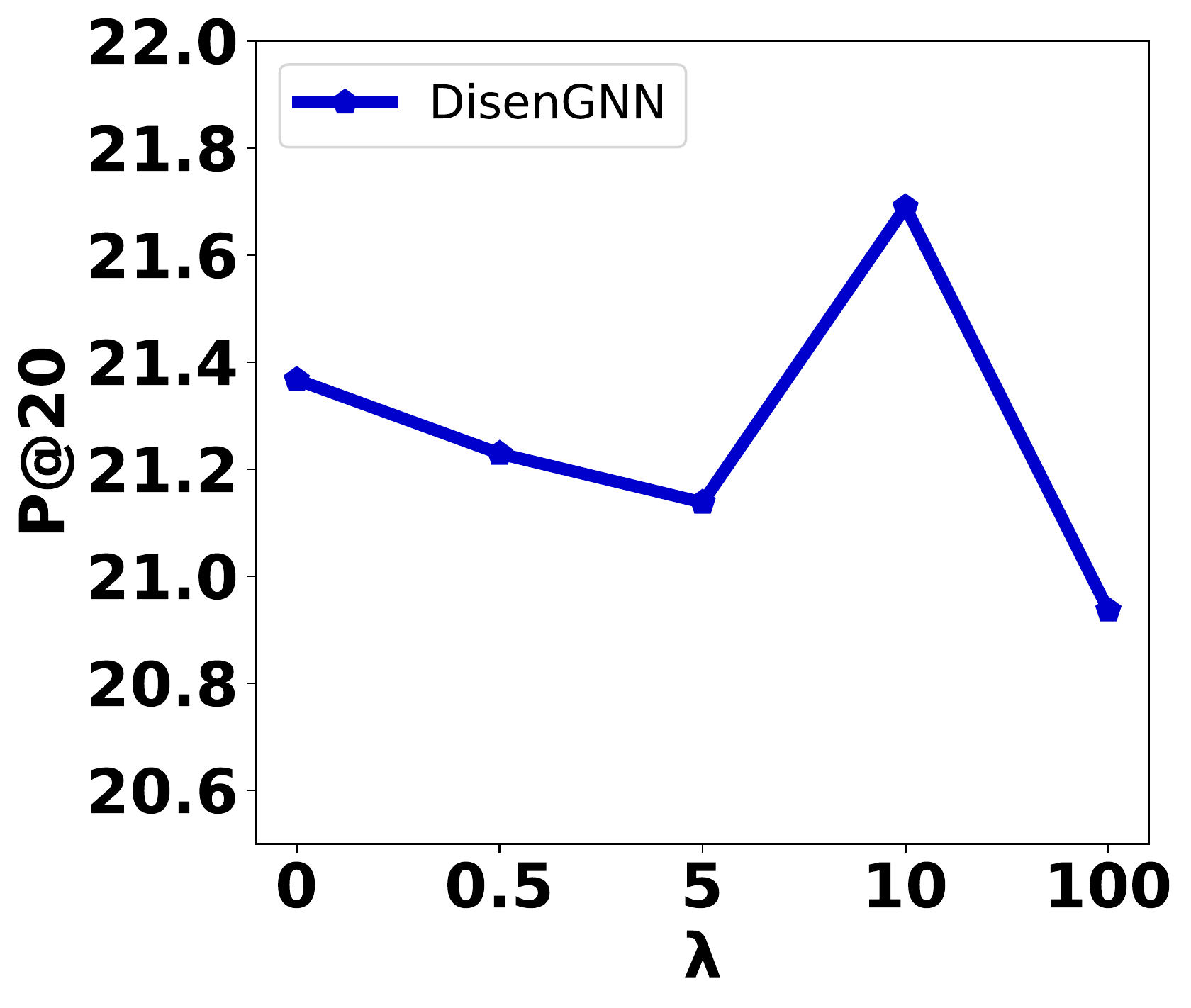}}
	\caption{Impact of regularization coefficient ($\lambda$).}
	\vspace{-10pt}
	\label{fig:cor}
\end{figure}

\textbf{Effectiveness of Residual Attention Mechanism. }
 We design a residual attention mechanism (RAM) module in DIEL layer to alleviate the over-smoothing problem. To evaluate its effectiveness, we conduct experiments  to compare Disen-GNN with a variant without using RAM (denoted as DisenGNN-w/o-RAM) in each DIEL layer. To save space, we only report the experiment results  on the Diginetica and Nowplaying datasets to show the effects of parameters (including Fig.~\ref{tab:ram}, Fig.~\ref{fig:feature} and Fig.~\ref{fig:cor}). 
 
 In this study, we evaluate the performance of Disen-GNN and DisenGNN-w/o-RAM with  different numbers of GGNN layers in each DIEL layer.  From Fig.~\ref{tab:ram}, we can see that Disen-GNN outperforms DisenGNN-w/o-RAM consistently across different numbers of GGNN layers by a large margin, demonstrating the positive effects of our RAM module on learning item embedding. In addition, with our RAM module, the performance is stable with the increasing of GGNN layers. In contrast, without RAM module, the performance drops sharply when the number of layers increased, especially on the Nowplaying dataset. The results show the capability of our RAM module on alleviating the over-smoothing problem. The best performance is achieved by stacking 2-3 layers. This is because although stacking more layers can exploit information from higher-order neighbors, there is noisy information in the information from neighbors far away, which negatively impacts the embedding learning~\cite{IMPGCN}.

\subsection{ Influence of key parameters (RQ4)}

 \textbf{Number of Factors ($K$).} 
 To study the effects of the factor numbers ($K$), we keep the item embedding size (i.e., $d$) unchanged, and tuning the value of $K$. We conduct experiments on the Diginetica and Nowplaying datasets and set $d$ to be 96 and 80 for the two datasets, respectively. Note that  the embedding size of factors is computed as $\frac{d}{K}$, which is varied with the change of $K$. The performance with varying factor numbers is shown in Fig. \ref{fig:feature}. In general, with the increase of factor number, the performance is generally increased till arriving a peak and then starts to drop. When $K=1$, the performance is worst, indicating that representing items with a holistic representation is insufficient to capture user purpose in a session. This also demonstrates the rationality of disentangling the item embedding and model the session purpose at the fine-granularity level. By comparing the performance of two datasets, we can find that the influence of factor numbers are different for them. This is because the two datasets are collected from different scenarios, and the user intents are driven by different factors for different scenarios. The best performance is obtained at different factor numbers for the two datasets, e.g., $K=5$ on the Diginetiva.  On the Nowplaying dataset, $P@20$ is still increasing when $K=12$ while MRR@20 drops when $K$ is larger than 6. This is possible as MRR takes the ranking position into consideration, while $P@20$ only considers whether the ground truth is at the top 20 results. The performance drops when the factor number is larger than a threshold. On the one hand, this suggests that too fine-grained factors may hurt the performance; on the other hand, because we fix the  item embedding size $d$, with a larger value of factor number, the embedding size for each factor becomes smaller, which may limit the expressiveness capability of the factor embedding (e.g., $\frac{d}{K} = 10$ when $K=8$ on Diginetiva).

\textbf{Regularization coefficient  ($\lambda$).} 
This final loss function consists of a prediction loss and an independent loss which aims to make factors independent. The regularization coefficient $\lambda$ is to control the weight of the independent loss in the training process. We tuned $\lambda$ in a wide range in each dataset for the best performance, as it is affected by the values of both losses. We use the recommendation performance based on $P@20$ on Nowplaying and Diginetica datasets to study its influence, since similar trends are observed based on $MRR@20$. The influence of $\lambda$ is shown in Fig.7. From the figure, we can see that a proper setting of $\lambda$ can significantly enhance the performance. When $\lambda=0$, the performance is relatively worse on both datasets, which indicates that making factors independent can effectively facilitate the embedding learning and thus enhance the performance. When $\lambda$ becomes too large, it weakens the guiding effects of prediction loss on the learning process, resulting in performance degradation. 

\section{ Conclusions }
In this paper, we  propose a novel Disentangled Graph Neural Network (Disen-GNN) model for session-based recommendation. Our model applies the disentangled representation learning to cast the item embedding into multiple chunks to represent different factors. The embedding of each factor is learned separately via GGNN with a factor-based similarity matrix, where each element represents the similarity between items based on a specific factor. In addition, distance correlation is adopted to enhance the independence between each pair of factors. By representing each item with independent factors, the session embedding is then generated by a weighted combination of all the items in a session with assigned factor-level attentions. Extensive experiments have been conducted on three benchmark datasets to evaluate the effectiveness of our model. Experimental results demonstrate the superiority of our model over existing methods consistently across all datasets. Moreover, ablation studies validate the validity of each component of our model. This work highlights the importance of considering the factor-level attention on modeling the anonymous user purpose in a session for SBR and also sheds light on the potential of applying disentangled learning to solve this problem. In the future,  we would like to explore the context information or the global information of all sessions to better capture user's interests on different factors of items for SBR.


\bibliographystyle{IEEEtran}
\bibliography{DisenGNN}
\end{document}